\documentclass[a4paper,fleqn]{cas-sc}

\usepackage[numbers,sort&compress]{natbib}
\usepackage[T1]{fontenc}
\usepackage{nicefrac}   % compact 1/2 etc.
\usepackage{microtype}  % microtypography
\usepackage{float}      % [H] placement in the appendices
\usepackage{url}        % \url for the data-availability link
\usepackage{cleveref}   % smart cross-referencing (loaded after the class' hyperref)
\definecolor{better}{HTML}{009E73}
\newcommand{\win}[1]{\textcolor{better}{#1}}
\newcommand{\mrk}[2][0.9]{\ensuremath{\mathord{\scalebox{#1}{$#2$}}}}
\Crefname{appendix}{}{}
\Crefname{appendix}{}{}

\def\tsc#1{\csdef{#1}{\textsc{\lowercase{#1}}\xspace}}
\tsc{WGM}
\tsc{QE}
\begin{document}
\let\WriteBookmarks\relax
\def\floatpagepagefraction{1}
\def\textpagefraction{.001}
%\linenumbers

\shorttitle{Manifold-adapted RBFs for reduced-order modelling of chaotic flows}

\shortauthors{M.~P\'erez Cuadrado et~al.}

\title [mode = title]{Manifold-adapted radial basis functions for reduced-order modelling of chaotic flows}




\author[1]{M.~P\'erez Cuadrado}%[<options>]

\cormark[1]


\ead{miguel.perez-cuadrado@citystgeorges.ac.uk}


% TODO (author): confirm CRediT roles.
\credit{Conceptualization, Methodology, Software, Formal analysis, Investigation, Writing -- original draft, Visualization}

\affiliation[1]{organization={Department of Engineering, City St George's, University of London},
            addressline={Northampton Square},
            city={London},
%          citysep={}, % Uncomment if no comma needed between city and postcode
            postcode={EC1V 0HB},
            country={United Kingdom}}

\author[1]{G.~M.~Cavallazzi}%[]

% Footnote of the second author

% Email id of the second author
%\ead{}


\credit{Methodology, Software, Writing -- review \& editing}

\author[1]{A.~Pinelli}%[]

\credit{Conceptualization, Supervision, Writing -- review \& editing, Funding acquisition}

%\affiliation[2]{organization={},
%%          citysep={}, % Uncomment if no comma needed between city and postcode

\cortext[1]{Corresponding author.}



\begin{abstract}
Chaotic systems often evolve on a low-dimensional attractor whose geometry varies from one
region to another. We propose a non-intrusive reduced-order model that reads this local
geometry by clustering and uses it to shape a radial basis library whose kernels adapt to
each region. Fitting the reduced velocity onto this library by one global least-squares solve
gives an explicit, differentiable vector field that reproduces the long-term statistics without
any use of the governing equations.
A radial basis field decays away from the data and cannot by itself return an escaped state.
The integration is therefore stabilised by a kinematic corrector, whose reported magnitude
measures how far each result rests on the learned field.
On Lorenz-63 the model recovers the attractor, its marginal densities and its Lyapunov spectrum.
On Lorenz-96 its valid prediction time matches typical configurations of neural-network and
reservoir-computing forecasters and trails their best-tuned ones, and the invariant measure is
reproduced on both the full state and on a reduced observable. On the Kuramoto--Sivashinsky equation and the
quasiperiodic Kolmogorov flow the model matches the energy distribution and spectrum of an
intrusive quantised-local Galerkin model and improves on a global Galerkin projection of the
same reduced dimension.
The recovery is dictated by the distance from a state to its nearest kernels, not by the
one-step regression error.
\end{abstract}


% TODO: required by Physica D — 3 to 5 bullets, max 85 characters each
% (including spaces); also uploaded as a separate file at submission.
%\item
%\item
%\item

%\nocite{*}

% Keywords
\begin{keywords}
reduced-order modelling \sep radial basis functions \sep chaotic dynamics \sep invariant measure \sep non-intrusive methods \sep data-driven modelling
\end{keywords}

\maketitle

\section{Introduction}
\label{sec:intro}

Control, forecasting and data assimilation all need a model that can be integrated forward in time
at low cost, often in real time \citep{brunton2015control, bauer2015, evensen2009}.
The systems concerned are often multi-scale and chaotic, among them turbulent flows
\citep{pope2000, frisch1995} and the atmosphere and the ocean \citep{vallis2017, kalnay2003}.
Building such models is hard for two reasons: the state is sensitive to small perturbations
\citep{lorenz1963, strogatz2015}, and an accurate simulation requires fine spatial and temporal
resolution \citep{moin1998, ishihara2009}, whose cost grows rapidly with the Reynolds number
\citep{choi2012, spalart2000}.
A high-fidelity simulation is therefore affordable once, and not many times.

A reduced-order model lowers this cost by evolving a small number of dominant coordinates in place
of the full state, at the price of a controlled loss of fidelity \citep{benner2015, rowley2017}.
However good, such a model can track the trajectory of a chaotic system only for a few Lyapunov
times, the inverse of the rate at which nearby trajectories diverge
\citep{benettin1980, lorenz1969, strogatz2015}.
Beyond this horizon the statistics remain well defined: the state visits the attractor according to
its physical invariant measure \citep{eckmann1985, young2002}, which is insensitive to the initial
condition and is read in practice from the energy spectrum, the probability densities of the state
and its correlations \citep{pope2000}.
These statistics are therefore the primary target
\citep{pathak2017, pathak2018, vlachas2020}, together with boundedness under long integration
\citep{schlegel2015}.
Short-term tracking, measured in Lyapunov times \citep{vlachas2018}, is a secondary test of the
pointwise accuracy of the field within the predictable window.

Such a model first needs a low-dimensional representation of the state.
The proper orthogonal decomposition (POD) supplies a linear basis from data \citep{sirovich1987},
the modes that capture the most variance, and autoencoders replace it with a nonlinear one that
compresses the state into fewer coordinates \citep{lee2020}.
When the governing equations are known, as the Navier--Stokes equations are in fluid mechanics, the
classical route projects them onto the POD basis and yields a Galerkin reduced-order model
\citep{holmes2012}.
\citet{colanera2025} refined this construction by partitioning the attractor into regions, fitting a
local basis within each, and switching between the bases as the state passes from one region to the
next.
Each local basis is tailored to its region, so fewer modes are needed and both the
reconstruction of the state and the long-term statistics improve.
Both constructions are intrusive: they require the governing equations, which are not always known.

Non-intrusive approaches remove this dependence on the governing equations by learning the reduced
model from the data alone.
Recurrent networks \citep{vlachas2020} and reservoir computers \citep{pathak2017, pathak2018}
advance the state accurately over short horizons and reproduce the long-term statistics of chaotic
systems.
They carry the dynamics in a high-dimensional internal state, however, and return no vector field
for the reduced coordinates.
Controller design and stability analysis need that field in explicit and differentiable form
\citep{rowley2017, brunton2015control}.
These methods predict the state without exposing its dynamics.

Non-intrusive methods that return this field divide by whether they assume the analytic
form of the dynamics in advance.
Sparse regression selects a few active terms from a library of candidate functions
\citep{brunton2016, loiseau2018}, operator inference fits reduced operators of a prescribed
polynomial form \citep{peherstorfer2016}, and spectral-submanifold reduction fits a polynomial
normal form on an invariant manifold anchored at a fixed point or periodic orbit
\citep{haller2016ssm, cenedese2022ssmlearn}, a construction extended to chaotic attractors by
\citet{liu2024chaoticssm}.
Once this form is fixed, the model is linear in its coefficients, which follow from a single
regression, but it is only as expressive as the candidate functions, operator form or polynomial
order assumed.

To avoid restricting the model to a particular form, the field can be built from a universal
approximator: a neural network, or a sum of local functions placed on the data, such as radial
basis functions (RBFs) \citep{broomhead1988, xiao2015}.
A neural ordinary differential equation \citep{chen2018}, whose vector field is a neural network, is
expressive, explicit and differentiable \citep{linot2022}.
Its training is expensive and non-convex, and its weights are global, so the field at a state is set
by the whole data set.
The model cannot tell how far a state lies from the data it was fitted on.

A sum of RBFs keeps these properties and avoids both drawbacks.
An RBF depends on the state only through its distance to a fixed centre.
We use Gaussian kernels, which decay rapidly away from the centre, and build the field as a weighted
sum of kernels centred on points of the data \citep{broomhead1988}.
RBFs have been used in this way to predict chaotic time series \citep{casdagli1989}, to advance POD
coefficients \citep{xiao2015}, and, with learned kernels, to learn dynamical systems
\citep{hamzi2021, darcy2025}.
Such a sum can approximate any continuous function on a compact set \citep{park1991, hartman1990},
so the field needs no assumed analytic form, and it is explicit and infinitely differentiable.
The weights enter the field linearly and follow from a single least-squares solve.
Each kernel is local, so the field at a state is set by the nearby data, and the distance to the
nearest centre measures how far that state lies from the data.

Kernels usually share one shape, fixed in advance or tuned globally
\citep{casdagli1989, xiao2015, hamzi2021}.
A chaotic attractor is not uniform: its geometry changes from one region to another
\citep{eckmann1985, ginelli2007}, so a single shape either overlaps across it and ill-conditions the
regression \citep{schaback1995}, or leaves gaps along it.
We instead adapt the kernels to the local geometry of the attractor, as oriented kernels allow
\citep{poggio1990, schwenker2001, wand1995, berry2016}.
The attractor is partitioned by clustering
\citep{lloyd1982, kaiser2014, fernex2021, colanera2025}, and the principal directions of each cluster
\citep{kambhatla1997} make every kernel wide along the attractor and narrow across it.
The clusters shape the kernels and nothing else.
The weights are fitted over the whole library in one solve, so the dynamics remain one continuous
and differentiable field, with no switching between regions.

Non-intrusive models are trained on states sampled on the attractor, so they do not learn the
dynamics that pull the state towards it.
Away from the data the RBF field decays and cannot return an escaped state, and networks fitted to
the same data drift off the attractor as well \citep{linot2022}.
Recovering the invariant measure requires a bounded trajectory.
The distance to the nearest centre tells when the state has left the data, so we add a kinematic
corrector that draws the state back only beyond a trust radius set from the training data.
The field on the attractor is unchanged, whereas trapping-region conditions
\citep{schlegel2015, kaptanoglu2021} constrain the whole state space.

The result is a non-intrusive reduced-order model for chaotic systems: a vector field built from
radial basis functions placed on the data and shaped by the local geometry of the attractor,
fitted by one regression, and integrated with a corrector that acts only where the state leaves
the data.
We assess it on the Lorenz-63 and Lorenz-96 systems, the chaotic Kuramoto--Sivashinsky equation
and the quasiperiodic Kolmogorov flow.

The recovery of the invariant measure is decided by how densely the kernels cover the attractor,
not by the accuracy with which the field fits the training velocities, and some of the most
accurate fits are among those that collapse.
The same coverage sets the contraction transverse to the attractor, so the Lyapunov spectrum
follows it as well.
Measured against other models, the field forecasts as far as typical configurations of recurrent
and reservoir networks and reproduces long-term statistics as well as an intrusive local Galerkin
model built from the governing equations.
\Cref{sec:method} describes the construction of the model, \Cref{sec:results} reports the
assessment on these systems, and \Cref{sec:conclusions} concludes.

\section{Method}
\label{sec:method}

\subsection{Reducing the coordinates}
\label{sec:method:coords}

We consider a dynamical system whose state evolves according to
\begin{equation}
	\dot{\mathbf{u}} = \mathbf{F}(\mathbf{u}), \qquad \mathbf{u}\in\mathbb{R}^{N},
	\label{eq:fom}
\end{equation}
where \(\mathbf{F}\) generates the dynamics.
For a partial differential equation, \(\mathbf{u}\) holds the field on a grid of \(N\) points, and
for a system of ordinary differential equations it holds the state variables.
The system is observed through \(M\) snapshots \(\mathbf{u}_m = \mathbf{u}(m\,\Delta t)\),
\(m = 1,\dots,M\), taken from a long trajectory on its attractor.
The interval \(\Delta t\) is a design parameter rather than the step of the full-order solver,
since it is the step at which the reduced model is later fitted and advanced, and it is reported
with each system in \Cref{sec:results}.

Low-dimensional coordinates follow from POD \citep{sirovich1987, holmes2012} applied to the
fluctuations about the temporal mean,
\begin{equation}
	\mathbf{u}'_m = \mathbf{u}_m - \bar{\mathbf{u}},
	\qquad
	\bar{\mathbf{u}} = \frac{1}{M}\sum_{m=1}^{M}\mathbf{u}_m ,
	\label{eq:mean}
\end{equation}
so that the mean state is carried exactly and no mode is spent describing it.
The fluctuations are collected as the columns of
\(\mathbf{X} = [\,\mathbf{u}'_1, \dots, \mathbf{u}'_M\,] \in \mathbb{R}^{N\times M}\), and the
singular value decomposition
\(\mathbf{X} = \boldsymbol{\Psi}\,\boldsymbol{\Sigma}\,\mathbf{Z}^{\top}\)
yields the POD modes \(\boldsymbol{\psi}_i\) as the columns of \(\boldsymbol{\Psi}\),
ordered by the singular values \(\sigma_i\), whose squares measure the variance each mode
captures.

The leading \(r\) modes are retained, with \(r\) the smallest number that captures a fraction
\(\varepsilon \in (0, 1]\) of the variance,
\begin{equation}
	\frac{\sum_{i=1}^{r}\sigma_i^2}{\sum_{i=1}^{\min(N,M)}\sigma_i^2} \ge \varepsilon .
	\label{eq:variance}
\end{equation}
The criterion is one of energy alone: it takes no account of how well the retained coordinates
close among themselves, which is a property of the dynamics and not of the basis.
Both \(\varepsilon\) and \(r\) are reported for each system in \Cref{sec:results}.
With the truncated basis \(\boldsymbol{\Psi}_r = [\boldsymbol{\psi}_1,\dots,\boldsymbol{\psi}_r]\),
the reduced coordinates and the reconstruction of the state are
\begin{equation}
	\mathbf{a}(t) = \boldsymbol{\Psi}_r^{\top}\big(\mathbf{u}(t)-\bar{\mathbf{u}}\big)
	\in \mathbb{R}^{r},
	\qquad
	\mathbf{u}(t) \approx \bar{\mathbf{u}} + \boldsymbol{\Psi}_r\,\mathbf{a}(t),
	\label{eq:reduced}
\end{equation}
and the projection of each snapshot gives the reduced trajectory
\(\mathbf{a}_m = \boldsymbol{\Psi}_r^{\top}\mathbf{u}'_m\) on which the rest of the method
operates.

We seek a closed model for these coordinates,
\begin{equation}
	\dot{\mathbf{a}} = \mathbf{f}(\mathbf{a}), \qquad
	\mathbf{f}:\mathbb{R}^{r}\to\mathbb{R}^{r}.
	\label{eq:rom}
\end{equation}
When the state is already low-dimensional, all modes are kept, \(\varepsilon = 1\), and the
method acts on the full state.
Writing the reduced velocity as a function of \(\mathbf{a}\) alone assumes that the retained
coordinates are self-contained, and the memory that truncation leaves behind sets a limit on what
any such field can reproduce.

\subsection{Clustering}
\label{sec:method:clustering}
A chaotic attractor is extended along some directions and compressed along others
\citep{eckmann1985}, and the orientation of these directions changes from one region to
another \citep{ginelli2007}.
No single kernel shape can follow that variation, so the attractor is partitioned into regions,
each with its own shape.
Within a region the data are taken as a local piece of a low-dimensional manifold: the leading
principal directions span an approximate tangent space, and the remaining directions its normal
complement.
Clustering finds these regions without supervision and at little cost, and a partition of this
kind exposes an organisation of complex flows that a single global description
misses \citep{kaiser2014, colanera2025}.
The partition does nothing beyond this: it fixes the kernel shapes of \Cref{sec:method:rbf} and
leaves the dynamics a single field fitted over the whole library.

\subsubsection{Clustering the reduced coordinates}
\label{sec:method:cluster:kmeans}
The clustering uses \(K\)-means \citep{lloyd1982, arthur2007}, which partitions the \(M\) points of the
reduced trajectory into \(K\) clusters, each represented by a centre
\(\mathbf{c}_k \in \mathbb{R}^r\), \(k = 1,\dots,K\).
Every point \(\mathbf{a}_m\) is assigned to the cluster whose centre is nearest in
Euclidean distance, and each centre is the mean of the points assigned to it,
\begin{equation}
	k(m) = \arg\min_{k}\,\lVert \mathbf{a}_m - \mathbf{c}_k \rVert,
	\qquad
	\mathbf{c}_k = \frac{1}{n_k} \sum_{m\,:\,k(m)=k} \mathbf{a}_m,
	\label{eq:kmeans}
\end{equation}
with \(n_k\) the number of points in cluster \(k\).
The centres and the assignment together minimise the total squared distance
of each point from its centre,
\begin{equation}
	\{\mathbf{c}_k\},\, k(\cdot) = \arg\min\,
	\sum_{m=1}^{M} \lVert \mathbf{a}_m - \mathbf{c}_{k(m)} \rVert^2 ,
	\label{eq:kmeans-obj}
\end{equation}
which \(K\)-means reduces by alternating the two steps of \Cref{eq:kmeans} from an initial set of
centres.
The partition is therefore a set of Voronoi cells in the reduced coordinates, with straight
boundaries and one isotropic metric, and it describes a thin and curved attractor only coarsely.
It is the least adaptive step in the construction.

We cluster the POD coordinates without rescaling.
The spread of each coordinate is proportional to its singular value, so the high-variance
modes dominate the Euclidean distance \citep{kaiser2014, burkardt2006}.
We assume that the high-variance modes are those that best
distinguish one region of the attractor from another.
Rescaling to unit variance would instead give the low-variance modes equal weight
(\Cref{fig:clustering}c,~d) \citep{fernex2021, colanera2025}.

\subsubsection{Resampling by arc length}
\label{sec:method:cluster:arc}
\begin{figure*}[pos=t]
	\centering
	\includegraphics[width=\textwidth]{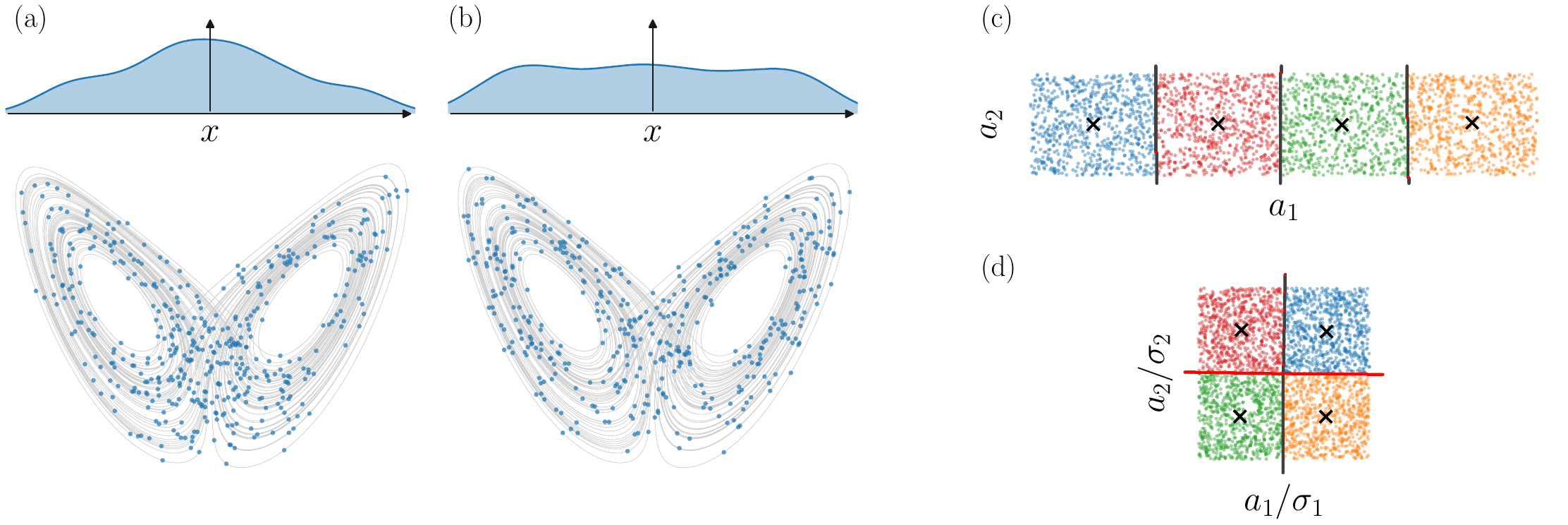}
	\caption{%
		Sampling and scaling in the clustering step.
		\emph{(a, b)}: the Lorenz-63 attractor sampled uniformly in time~(a) and in
		arc length~(b), with the marginal density of \(x\) above each panel.
		\emph{(c, d)}: clustering on the raw~(c) and unit-variance~(d) coordinates.
	}
	\label{fig:clustering}
\end{figure*}
Sampling uniformly in time places the points with a density inversely proportional to the speed \citep{eckmann1985, young2002}.
\Cref{eq:kmeans-obj} sums over the points, so the centres concentrate where the state
lingers \citep{zador1982, gersho1979, graf2000}, and the partition reflects the dwell time
rather than the geometry of the attractor, leaving the fast regions covered by few clusters.

The trajectory is therefore resampled uniformly in its arc length,
\begin{equation}
	s(t) = \int_0^t \lVert \dot{\mathbf{a}}(\tau) \rVert \, \mathrm{d}\tau ,
	\label{eq:arclength}
\end{equation}
the distance travelled in the reduced space.
The velocity \(\dot{\mathbf{a}}\) is estimated by a five-point finite-difference stencil on the
full-order record at its own step, and \(s\) by the trapezoidal rule.
The total length is divided into \(M-1\) equal intervals, and the reduced state at the
corresponding times is obtained by linear interpolation.
The resampled points are equispaced along the attractor, as \Cref{fig:clustering}(a, b) shows, and
the clustering of \Cref{sec:method:cluster:kmeans} acts on them.
When the speed is nearly uniform, \(s\) is proportional to \(t\) and the resampling may be
omitted.
The resampled points serve the clustering and the cluster covariance of \Cref{eq:localcov}, which
fixes the kernel shapes.
The regression of \Cref{sec:method:conditioning} uses the original snapshots.

\subsubsection{Choosing the number of clusters}
\label{sec:method:cluster:K}
The number of clusters \(K\) sets how finely the attractor is partitioned, and no single
value suits every system.
More clusters give smaller regions with simpler dynamics, but each holds fewer snapshots and
differs less from its neighbours.
The choice of \(K\) therefore takes two steps: the Bayesian information criterion (BIC) gives a range of
plausible values, and a comparison of neighbouring clusters selects among them.

The range comes from the BIC \citep{schwarz1978, pelleg2000},
\begin{equation}
	\mathrm{BIC}(K) = M \log J_K + K \log M
	- \frac{2}{r} \sum_{k=1}^{K} n_k \log \frac{n_k}{M},
	\label{eq:bic}
\end{equation}
where \(J_K\), the distortion, is the minimum of \Cref{eq:kmeans-obj} at \(K\) clusters.
The first term falls as clusters are added, while the last two penalise their number and
the spread of the points among them, with the normalisation by \(r\) following
\citet{colanera2025}.
The plausible values of \(K\) lie at and just beyond the shoulder of \(\mathrm{BIC}(K)\), its point
of largest curvature \citep{colanera2025}, past which each further cluster lowers the criterion by
less.
On a continuous attractor the distortion does not saturate as clusters are added, so
\(\mathrm{BIC}(K)\) need not have a minimum and its shoulder can be faint.

The criterion marks where further clusters stop adding information, but it says nothing
about whether the resulting regions differ in shape or in dynamics, which is what motivates
partitioning the reduced space.
Neighbouring clusters, those between which the state passes directly, are therefore compared,
first in shape and then in dynamics.

The shape of cluster \(k\) is described by the covariance of its centred coordinates
\citep{kambhatla1997},
\begin{equation}
	\mathbf{C}_k = \frac{1}{n_k} \sum_{m\,:\,k(m)=k}
	(\mathbf{a}_m - \mathbf{c}_k)(\mathbf{a}_m - \mathbf{c}_k)^{\top}
	= \mathbf{V}_k \boldsymbol{\Lambda}_k \mathbf{V}_k^{\top},
	\label{eq:localcov}
\end{equation}
whose eigenvectors \(\mathbf{v}_{k,p}\) and eigenvalues
\(\lambda_{k,1} \ge \dots \ge \lambda_{k,r}\) give its principal directions and their variances.
The same geometry shapes the kernels of \Cref{sec:method:dynamics}.

Two statistics compare neighbours \(i\) and \(j\).
The orientation is the largest principal angle between their leading directions
\citep{bjorck1973},
\begin{equation}
	\sin\theta_{\max}^{ij} = \sqrt{1 - s_{\min}^2\big(\widehat{\mathbf{V}}_i^{\top}\widehat{\mathbf{V}}_j\big)}
	\in [0, 1],
	\label{eq:angle}
\end{equation}
where \(\widehat{\mathbf{V}}_k\) collects the leading directions that capture \(99\%\) of the
variance of cluster \(k\), both truncated to the smaller of the two counts, and
\(s_{\min}\) denotes the smallest singular value.
The angle vanishes when the two subspaces coincide.
The elongation is the participation ratio \citep{bell1970},
\begin{equation}
	P_k = \frac{\big(\sum_{p=1}^{r} \lambda_{k,p}\big)^2}{\sum_{p=1}^{r} \lambda_{k,p}^2}
	\in [1, r],
	\label{eq:participation}
\end{equation}
the effective number of occupied directions, compared through \(|P_i - P_j|\).
Both are small where the partition only slices a smooth manifold.

The covariance of the arc-resampled points is blind to how fast the state moves, so each
cluster is also summarised by its characteristic speed
\begin{equation}
	v_k = \frac{1}{n_k} \sum_{m\,:\,k(m)=k} \lVert \dot{\mathbf{a}}_m \rVert ,
	\label{eq:speed}
\end{equation}
with \(\dot{\mathbf{a}}_m\) the velocity of \Cref{sec:method:cluster:arc} at the resampled
points, so that \(v_k\) is the mean speed along the arc rather than in time.
Neighbouring regions are dynamically distinct when the relative speed gap
\(|v_i - v_j| / (v_i + v_j) \in [0, 1]\) is appreciable.
Within the range the BIC allows, \(K\) is taken as the smallest partition whose neighbouring
clusters differ in both orientation and speed.
\subsection{Reduced dynamics}
\label{sec:method:dynamics}

\begin{figure*}[pos=t]
	\centering
	\includegraphics[width=\textwidth]{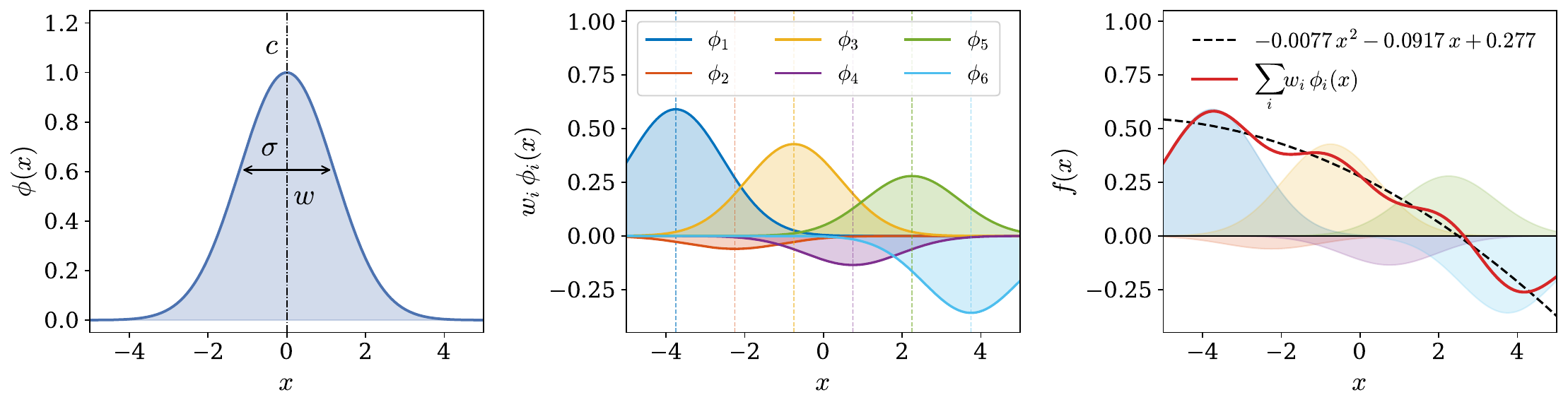}
	\caption{%
		The radial basis library in one dimension.
		\emph{Left:} a single kernel \(\phi\), fixed by its centre \(c\) and width
		\(\sigma\), and scaled by a weight \(w\).
		\emph{Centre:} several kernels at different centres.
		\emph{Right:} weighted sum \(\sum_i w_i \phi_i(x)\) (solid) approximating the
		target (dashed).%
	}
	\label{fig:rbf}
\end{figure*}

The field \(\mathbf{f}\) of \Cref{eq:rom} carries the dynamics in the reduced space.
We assume no functional form for it and build it from a library of \(n_c\) kernels \(\phi_i\)
anchored at points of the reduced trajectory, each appreciable only in its own neighbourhood,
combined with weights \(\boldsymbol{\xi}_i\),
\begin{equation}
	\dot{\mathbf{a}} = \mathbf{f}(\mathbf{a}) \approx
	\sum_{i=1}^{n_c} \boldsymbol{\xi}_i \, \phi_i(\mathbf{a}).
	\label{eq:rbf-model}
\end{equation}
The field at a point is then set by the data around it, and the partition of
\Cref{sec:method:clustering} enters through the \(\phi_i\), each taking the shape of the region it
sits in.
Locality is bought at a price: the field decays where there are no centres, so it cannot carry a
state that has left the data, and \Cref{sec:method:integration} returns to this.
\Cref{fig:rbf} shows the construction in one dimension: each centre contributes a weighted kernel
and their sum follows the target.
Three choices remain: the positions of the centres, the shapes of the kernels and the weights.
The first two follow the local geometry of the data (\Cref{sec:method:rbf}) and the third comes
from a regression against the measured velocities (\Cref{sec:method:conditioning}).

\subsubsection{Radial basis library and anisotropic kernels}
\label{sec:method:rbf}
The kernels are Gaussian RBFs \citep{micchelli1986, powell1987, broomhead1988},
\begin{equation}
	\phi(\mathbf{a}) = \exp\!\left(
	-\frac{\lVert \mathbf{a} - \mathbf{c} \rVert^2}{2\sigma^2}
	\right),
	\label{eq:gaussian-rbf}
\end{equation}
peaked at the centre \(\mathbf{c}\) and negligible beyond a few multiples of the width
\(\sigma\).
Two properties make this library fit the task: it is infinitely differentiable, so the field of
\Cref{eq:rbf-model} is smooth enough to integrate, and it is local, so the field near a point is
carried by the centres around it.
Locality does not restrict what the library can represent \citep{park1991, hartman1990}.
A global budget of \(n_c\) centres is divided equally among the \(K\) clusters rather than in
proportion to the number of snapshots each holds, so that every region is covered at the same
budget whatever its population.
Within each cluster the centres are placed by farthest-point sampling \citep{gonzalez1985},
starting from the point farthest from the centroid and adding at each step the cluster point
farthest from those already chosen.
This gives a roughly uniform cover of every region, with neither large gaps nor clumping \citep{demarchi2005}.

The data cloud is anisotropic, elongated along the high-variance directions and squeezed along the
low-variance ones (\Cref{fig:clustering}c,~d).
An isotropic Gaussian suits neither direction.
Wide enough for the elongated directions, it covers states the system never visits.
Narrow enough for the squeezed ones, it overfits.
The orientation and spread change across the attractor, so the kernel shape has to be set
region by region.

We resolve this mismatch by giving each kernel its own anisotropy, replacing the scaled
Euclidean distance of \Cref{eq:gaussian-rbf} with a Mahalanobis distance
\citep{mahalanobis1936} under a local metric \(\mathbf{W}_k\), as in hyper basis function
networks \citep{poggio1990, schwenker2001},
\begin{equation}
	\phi_i(\mathbf{a}) = \exp\!\left(
	-\tfrac{1}{2}\,
	(\mathbf{a}-\mathbf{c}_i)^{\top}\,
	\mathbf{W}_{k(i)}\,
	(\mathbf{a}-\mathbf{c}_i)
	\right),
	\label{eq:aniso-rbf}
\end{equation}
where the centre \(\mathbf{c}_i\) belongs to cluster \(k(i)\).

This metric is taken from the local covariance of \Cref{eq:localcov}, scaled by a single
multiplier \(\alpha\) common to all clusters,
\begin{equation}
	\mathbf{W}_k = \big(\alpha^{2}\,\mathbf{C}_k\big)^{-1},
	\label{eq:metric}
\end{equation}
under which distance is large along the directions where the cluster is narrow and small along
those where it is broad.
Each kernel thus inherits the principal axes of its parent cluster, with width
\(\alpha\sqrt{\lambda_{k,p}}\) along axis \(p\).
The multiplier \(\alpha\) sets the overlap between neighbouring kernels.
Together with the budget \(n_c\), \(\alpha\) is tuned for each system (\Cref{sec:results}):
wider kernels bridge the gaps between centres, but overlap more and ill-condition the regression
of \Cref{sec:method:conditioning}.
Neither is chosen by minimising the error of the fit, which \Cref{sec:results} shows does not
decide whether the integrated model recovers the invariant measure.

\subsubsection{Regression}
\label{sec:method:conditioning}

The centres and widths are fixed, so the kernels are known functions of \(\mathbf{a}\) and only
the weights of \Cref{eq:rbf-model} remain to be found.
With the kernel values at \(\mathbf{a}\) collected in a row and the weights in a matrix
\(\boldsymbol{\Xi} = [\boldsymbol{\xi}_1, \dots, \boldsymbol{\xi}_{n_c}]^{\top}\), one column per
reduced coordinate, the model reads
\begin{equation}
	\mathbf{f}(\mathbf{a}) = \mathbf{G}(\mathbf{a})\,\boldsymbol{\Xi},
	\qquad
	\mathbf{G}(\mathbf{a}) = \big[\phi_1(\mathbf{a}), \dots, \phi_{n_c}(\mathbf{a})\big],
	\label{eq:rbf-matrix}
\end{equation}
which is linear in the weights.
The weights are computed by least squares \citep{broomhead1988}, through the singular value
decomposition of \(\mathbf{G}\) \citep{golub1970}.
No penalty is placed on the weights: the conditioning is controlled by \(\alpha\) and \(n_c\)
rather than by regularising the solve.
Although each kernel carries the anisotropy of its parent cluster, the regression is global: all
the weights are found at once, and \(\mathbf{f}\) is a single smooth field on the whole reduced
space.
The kernels overlap, so the columns of \(\mathbf{G}\) are nearly collinear and only the
combination \(\mathbf{G}(\mathbf{a})\,\boldsymbol{\Xi}\) is determined.
The model is therefore read as a vector field and not as a list of coefficients.

\subsection{Time integration and the kinematic corrector}
\label{sec:method:integration}

The fitted model, \Cref{eq:rbf-model}, is integrated with the classical fourth-order
Runge--Kutta scheme \citep{hairer1993} at a fixed step equal to the snapshot spacing.
Recovering long-term statistics requires the prediction to remain bounded, which data-driven
models do not guarantee unless boundedness is built into them \citep{kaptanoglu2021}.
Small errors accumulate and can carry the state out of the region sampled by the training data,
where the model has no support and the prediction drifts off the attractor or diverges.
This occurs whether the model advances the state with a recurrent map \citep{vlachas2020} or
with a learned vector field \citep{linot2022, linot2023}, and such models cannot tell when it has
happened, so a correction is hard to target.

Here an escape is detectable, since the kernels that support the field are local and centred on
the data.
The distance from a state to its nearest centre, measured in the metric of that centre's kernel,
\begin{equation}
	d(\mathbf{a}) = \min_{i} \big\lVert \mathbf{W}_{k(i)}^{1/2}(\mathbf{a} - \mathbf{c}_i) \big\rVert,
	\label{eq:dist}
\end{equation}
tells at every step how far the state lies from the data.

A corrector \(\mathbf{g}(\mathbf{a})\) added to the learned field then returns the state towards
the data,
\begin{equation}
	\dot{\mathbf{a}} = \mathbf{f}(\mathbf{a}) + \mathbf{g}(\mathbf{a}).
	\label{eq:rom-corrected}
\end{equation}
The corrector should vanish where the data lie, so that the learned field is integrated unchanged
there, switch on gradually rather than abruptly, and pull the state back towards the centres with
a strength that grows with its distance from them.
These requirements divide the space around each cluster into three regions.
They are set by a trust radius \(R_k\), the \(q\)-quantile of \Cref{eq:dist} over the snapshots of
cluster \(k\), with \(q\) close to one, and by a fraction \(\eta \in (0, 1)\) of \(R_k\) that bounds
the inactive region,
\begin{equation}
	\mathbf{g}(\mathbf{a}) =
	\begin{cases}
		\mathbf{0}, & d(\mathbf{a}) \le \eta R_k, \\[4pt]
		-\,\gamma \left( \dfrac{d(\mathbf{a}) - \eta R_k}{R_k\,(1 - \eta)} \right)^{\mu}
		\hat{\mathbf{n}}(\mathbf{a}), & d(\mathbf{a}) > \eta R_k,
	\end{cases}
	\label{eq:corrector-term}
\end{equation}
where \(k\) is the cluster of the nearest centre in \Cref{eq:dist} and \(\hat{\mathbf{n}}\) is the
outward unit vector from that centre to the state, so that the pull points back towards the data.
Within \(\eta R_k\) the corrector is inactive.
Between \(\eta R_k\) and \(R_k\) it rises continuously from zero to its nominal strength
\(\gamma\), and beyond \(R_k\), outside the data, it keeps growing with the distance.
The fraction \(\eta\) sets the width of this transition, the exponent \(\mu\) its shape and
\(\gamma\) the intensity of the pull.
\begin{figure}[pos=H]
	\centering
	\includegraphics[width=.9\textwidth]{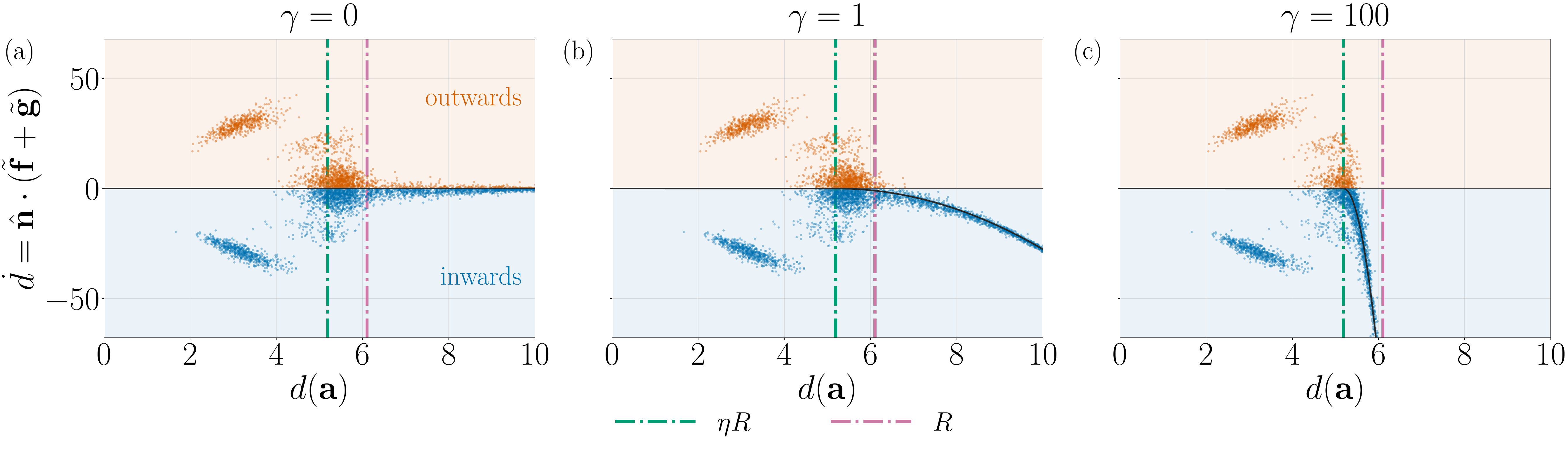}
	\caption{%
		The radial velocity against the distance $d$, at strengths \emph{(a)}~$\gamma = 0$, 
		\emph{(b)}~$\gamma = 1$ and \emph{(c)}~$\gamma = 100$ for the Lorenz-96 fit at $F = 8$ and $r = 40$, with $K = 10$ clusters and $n_c = 4000$ . 
	}
	\label{fig:balance}
\end{figure}
Three of these four parameters carry no units.
The quantile \(q\) reads the trust radius off the distribution of \Cref{eq:dist} over the snapshots
of each cluster, shown in \Cref{fig:app:calibration}(a),
\begin{equation}
	R_k = Q_q\big( d(\mathbf{a}) \big), \qquad q = 0.99,
	\label{eq:Rk}
\end{equation}
so the field is trusted as far out as the training data reach, and \(q\) sets how many snapshots
are left outside.
The fraction \(\eta\) and the exponent \(\mu\) shape the transition, how early the pull comes on
and how smoothly it rises.
Both may be moved from one case to another.
We hold them at \(\eta = 0.85\) and \(\mu = 2\) throughout.

The strength \(\gamma\) is the exception.
It is the only one of the four with units, a velocity in the reduced coordinates, so it has to be
measured against the field it corrects rather than fixed once for every case.
The distance supplies the reading.
The squared distance is taken as a Lyapunov candidate \citep{khalil2002},
\begin{equation}
	V(\mathbf{a}) = d(\mathbf{a})^2 ,
	\label{eq:lyapunov}
\end{equation}
whose sublevel set \(\{V \le R_k^2\}\) is the trust region itself.
Its rate of change along the flow,
\begin{equation}
	\dot V = 2\,d\,\dot d ,
	\qquad
	\dot d = \hat{\mathbf{n}} \cdot \big( \tilde{\mathbf{f}} + \tilde{\mathbf{g}} \big),
	\label{eq:vdot}
\end{equation}
with \(\tilde{\mathbf{f}} + \tilde{\mathbf{g}}\) the integrated field in the whitened coordinates
in which \(d\) is measured, is negative where the field draws the state back towards the data and
positive where it drives the state away.

\Cref{fig:balance} reads \(\dot d\) against \(d\) for a fitted model, with the onset \(\eta R_k\)
and the trust radius \(R_k\) marked.
Without correction, panel~(a), the field decays to zero beyond the data and leaves a set of states
with \(\dot V > 0\) past \(R_k\).
These are the paths along which a trajectory escapes, and they follow from a basis that decays
away from the data rather than from any one fit.
As the strength grows those states cross into \(\dot V < 0\), panel~(b).
Too large a strength turns the states inside the buffer as well, and so alters the dynamics on the
attractor, panel~(c).

The corrector is correctly sized when the inward pull beyond \(\eta R_k\) outweighs the outward
drift the field leaves there, and no more.
That outward drift is measurable, so \(\gamma\) need not be tuned by hand.
In the shell \(\{d > \eta R_k\}\) where the corrector acts, the strength matches the drift that
remains,
\begin{equation}
	\gamma = s\, P^{+}_{95}\big( \hat{\mathbf{n}} \cdot \tilde{\mathbf{r}} \big),
	\qquad
	\tilde{\mathbf{r}} = \tilde{\mathbf{v}} - \tilde{\mathbf{f}},
	\label{eq:kk}
\end{equation}
the ninety-fifth percentile of the outward residual velocity across the shell, with a safety
factor \(s = 1.5\), so that the pull exceeds the drift it has to cancel.
Here \(\tilde{\mathbf{r}}\) is the residual the field leaves behind, the empirical rate
\(\tilde{\mathbf{v}}\) minus the learned field \(\tilde{\mathbf{f}}\).
It is this residual, and not the full velocity, that sets the scale: the corrector should cancel
what the field gets wrong, and never oppose the dynamics the field already carries.

\Cref{app:corrector} runs the calibration on Lorenz-96 and shows there that the guarded model keeps
the long-term statistics of the full-order system, and that the corrector acts only during rare
excursions.
The magnitude of the correction relative to the reduced velocity is reported in \Cref{app:extra}
for the Lorenz-96, Kuramoto--Sivashinsky and Kolmogorov cases.

The corrector is a design choice that follows from the data.
The snapshots come from a trajectory on the attractor, as when a physical system is measured but
cannot be perturbed, so they carry no information on how the state returns to it.
With full control of the system, trajectories could instead be started away from the attractor
and their transients added to the training set, so that the fitted field itself learns the return.
Such control is rarely available, and the corrector supplies the return without it.

\clearpage
\section{Results}

\label{sec:results}
The model is assessed on four systems, each posing a different question.
The Lorenz-63 system has an attractor, long-term statistics and Lyapunov spectrum known to high
accuracy, so the learned field can be checked against ground truth at every stage.
It is the case in which each ingredient of the method, the partition, the kernel widths and the
corrector, can be varied on its own.

The Lorenz-96 system asks how the model compares with other non-intrusive, data-driven models,
the recurrent networks and reservoir computers of \citet{vlachas2020}, when it acts on the full
state and on a reduced set of coordinates.

The Kuramoto--Sivashinsky equation and the Kolmogorov flow ask how it compares with intrusive
models that have access to the governing equations, the Galerkin and quantised-local models of
\citet{colanera2025}.

For every system, the full-order model is integrated with a step $h$ and sampled every
$\Delta t$ time units, the interval at which the reduced model is fitted and advanced
(\Cref{sec:method:integration}), and both are given with each case.
The velocities $\dot{\mathbf a}_m$ that the regression targets are computed by the five-point
stencil of \Cref{sec:method:cluster:arc} on the full-order record at its own step, before the
record is thinned to $\Delta t$.
The first $80\%$ of the record is used to build the model and the remaining $20\%$ is held
out for testing.
The invariant measure is the primary target throughout and short-term forecasting a secondary one
(\Cref{sec:intro}), so a model that follows a particular trajectory for longer is not for that
reason the better model.

\subsection{Lorenz-63}
\label{sec:results:l63}

A field fitted accurately to the measured velocities does not imply that it will reproduce the attractor once it is
integrated, and Lorenz-63 \citep{lorenz1963, strogatz2015} is where the two can be separated.
It is a canonical benchmark on which data-driven and reduced-order models of chaotic dynamics are
calibrated and tested \citep{brunton2016, vlachas2018, pathak2017}.
The system reads
\begin{equation}
	\frac{\mathrm{d} x}{\mathrm{d} t} = \sigma (y - x), \qquad
	\frac{\mathrm{d} y}{\mathrm{d} t} = x (\rho - z) - y, \qquad
	\frac{\mathrm{d} z}{\mathrm{d} t} = x y - \beta z,
	\label{eq:lorenz63}
\end{equation}
with the classical parameters $\sigma = 10$, $\rho = 28$ and $\beta = 8/3$.
It is integrated with a step $h = 0.005$ and sampled every $\Delta t = 0.01$ time
units, over $500$ time units.

\begin{figure}[pos=H]
	\centering
	\includegraphics[width=\textwidth]{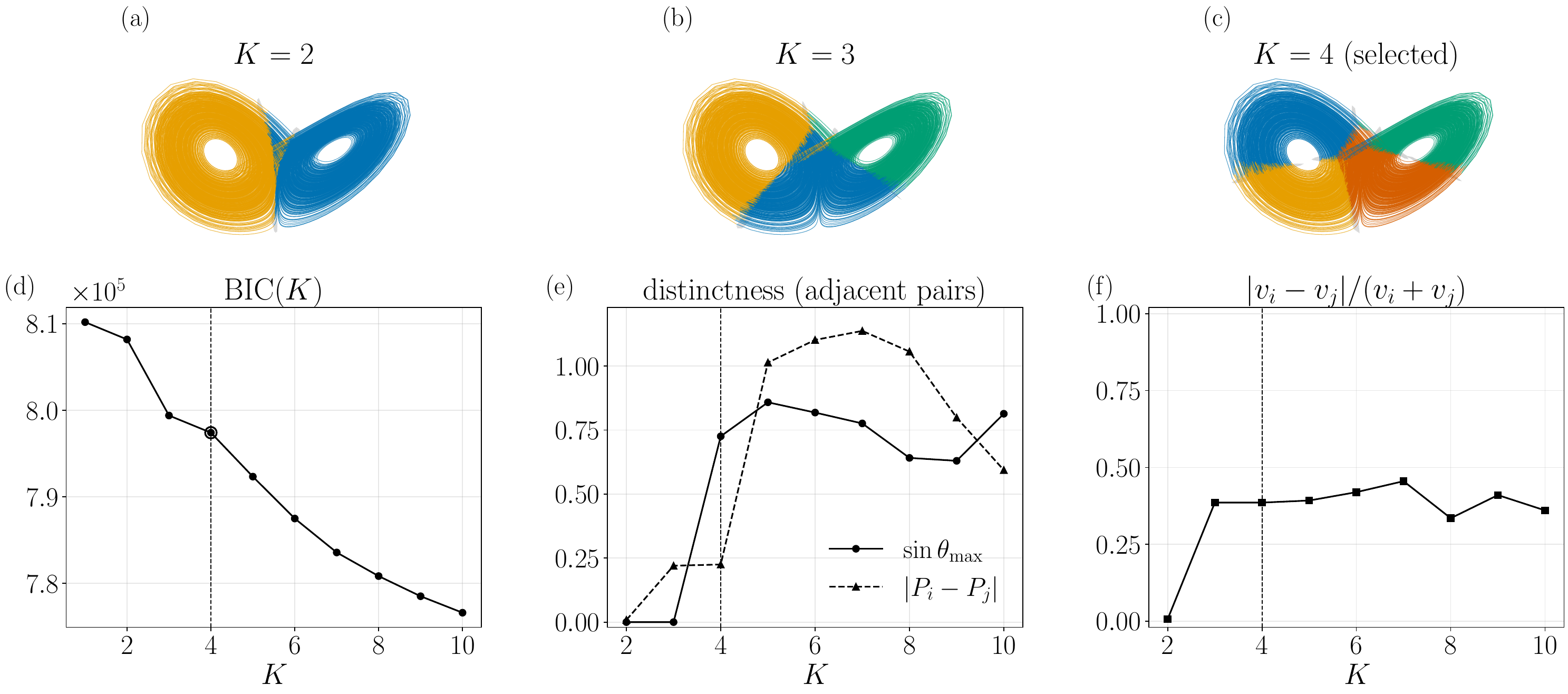}
	\caption{%
		Clustering of the Lorenz-63 state and selection of the number of regions.
		\emph{(a--c)} the attractor partitioned by $K$-means at $K = 2, 3, 4$, coloured by cluster.
		\emph{(d)} the Bayesian information criterion. \emph{(e)} the principal angle between neighbouring clusters.
		\emph{(f)} the relative speed gap between neighbouring clusters.
	}
	\label{fig:l63clusters}
\end{figure}

The test assesses only how well the method learns the dynamics, so the state is not reduced:
the field is fitted on $(x, y, z)$ itself, the case $\varepsilon = 1$ of
\Cref{sec:method:coords}.Following \Cref{sec:method:clustering}, the three-dimensional trajectory is resampled and clustered
at $K = 2, 3, 4$, with the selection metrics carried up to $K = 10$ (\Cref{fig:l63clusters}).

The Lorenz attractor is a continuous set and not a collection of separated groups, so the
distortion $J_K$ of \Cref{eq:kmeans-obj} does not saturate as clusters are added
\citep{zador1982, graf2000}, and the information criterion has no minimum
(\Cref{fig:l63clusters}d).
Its shoulder, near $K = 4$, is faint, so the criterion alone does not fix $K$, and the choice is
guided by the shape and dynamics indicators of \Cref{sec:method:cluster:K}.

The speed gap (\Cref{fig:l63clusters}f) responds first: it jumps to $0.38$ at $K = 3$, where a cluster isolates the slow
region near the saddle between the lobes, and stays flat beyond.
Up to $K = 3$ every cluster spans the curvature of the attractor and keeps all three directions, so
the principal angle vanishes.
At $K = 4$ two of the clusters are flat enough to lose a direction, and the angle rises to about
$0.7$ (\Cref{fig:l63clusters}c,~e).
$K = 4$ is therefore the smallest partition whose neighbouring clusters differ in both orientation
and speed, and the criteria of \Cref{sec:method:cluster:K} select it.
The results below are nonetheless shown for $K = 1$ to $4$, to expose the effect of the clustering
on the model.

The regression of \Cref{eq:rbf-matrix} is solved for widths $\alpha$ from $0.5$ to $5$, budgets
$n_c$ from $25$ to $800$ and $K$ from $1$ to $4$.
The quality of each fit is measured by its one-step root-mean-square error on the target
$\dot{\mathbf a}$, normalised by the standard deviation of $\dot{\mathbf a}$.
\begin{figure}[pos=H]
	\centering
	\includegraphics[width=\textwidth]{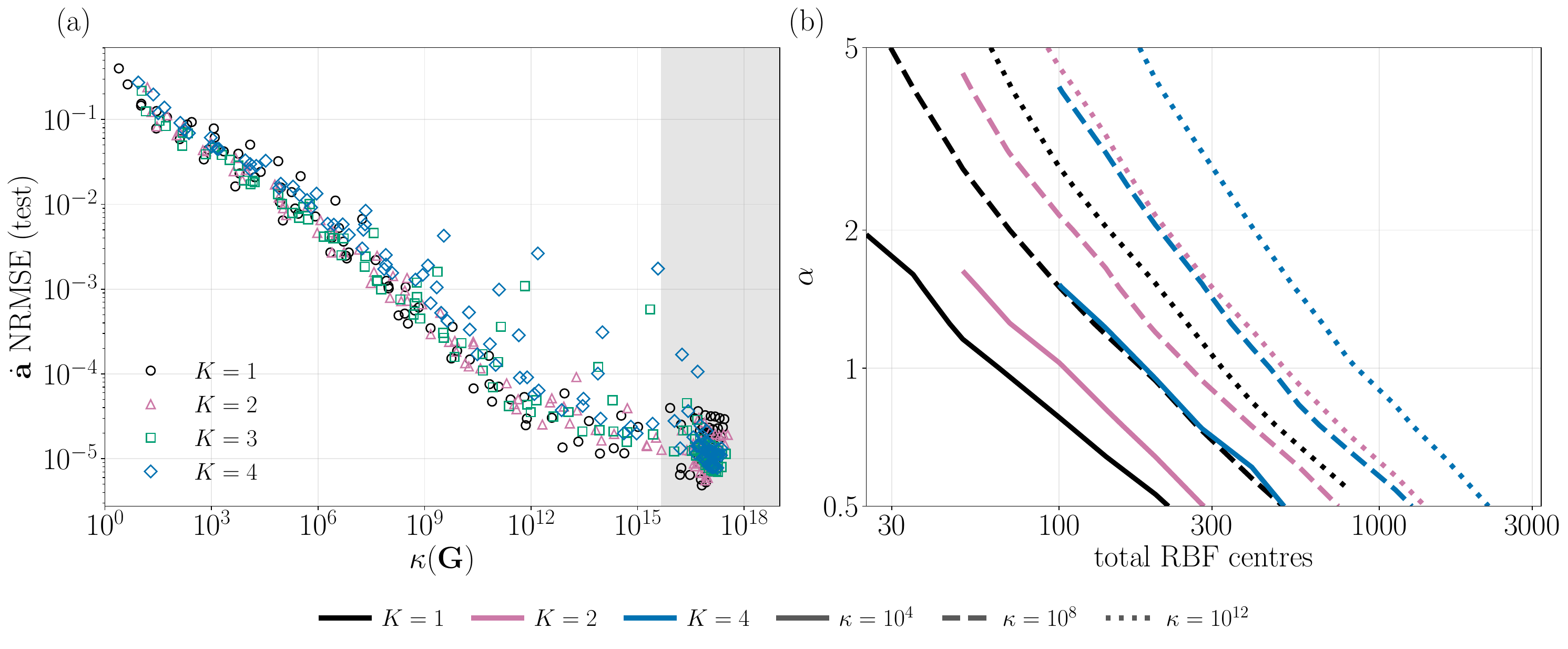}
	\caption{%
		A-priori regression on Lorenz-63. \emph{(a)} test NRMSE of $\dot{\mathbf a}$ against the 
		condition number $\kappa(\mathbf{G})$ of the library, for $K = 1$ to $4$ across widths and numbers of kernels. 
		\emph{(b)} contours $\kappa = 10^4, 10^8, 10^{12}$ in the plane of total kernels $n_c$ and width $\alpha$, 
		for $K = 1, 2, 4$.
	}
	\label{fig:l63conditioning}
\end{figure}
The error of the fit is set by the condition number $\kappa(\mathbf{G})$ (\Cref{eq:rbf-matrix}) of the
library (\Cref{fig:l63conditioning}a).
The fits fall on one curve, along which the error decreases as $\kappa$ grows, until round-off
saturates it near $\kappa \approx 10^{16}$.
For this metric $\alpha$ and $K$ matter only through $\kappa$, apart from a few fits at $K = 3$ and
$4$ that lie above the curve.

A given $\kappa$ is reached by many pairs of width and kernel count, from many narrow kernels to few
wide ones (\Cref{fig:l63conditioning}b), so a target $\kappa$ leaves the choice of $\alpha$ and $n_c$
open.
At equal width and total budget, partitioning lowers $\kappa$ by several decades, since each kernel
scales with its cluster and not with the whole attractor.
A target $\kappa$ is then reached with more kernels, that is, with a denser cover of the attractor,
whose effect only the integration shows.

All the fitted models are integrated for $500$ time units, about $450$ Lyapunov times.
The trust radii of the corrector are recalibrated for each model, with its other parameters
fixed (\Cref{sec:method:integration}).

The recovery of the invariant measure is scored by the Wasserstein-1 distance between the marginal
densities of the model and of the reference, summed over the coordinates,

\begin{equation}
	W_1 = \sum_{i=1}^{r} \int \big| F_i^{\mathrm{ROM}}(s) - F_i^{\mathrm{FOM}}(s) \big| \, \mathrm{d}s,
	\label{eq:w1}
\end{equation}
with $F_i$ the cumulative distribution of $a_i$ \citep{villani2009}, and its median over
five initial conditions is reported.

\begin{figure}[pos=H]
	\centering
	\includegraphics[width=\textwidth]{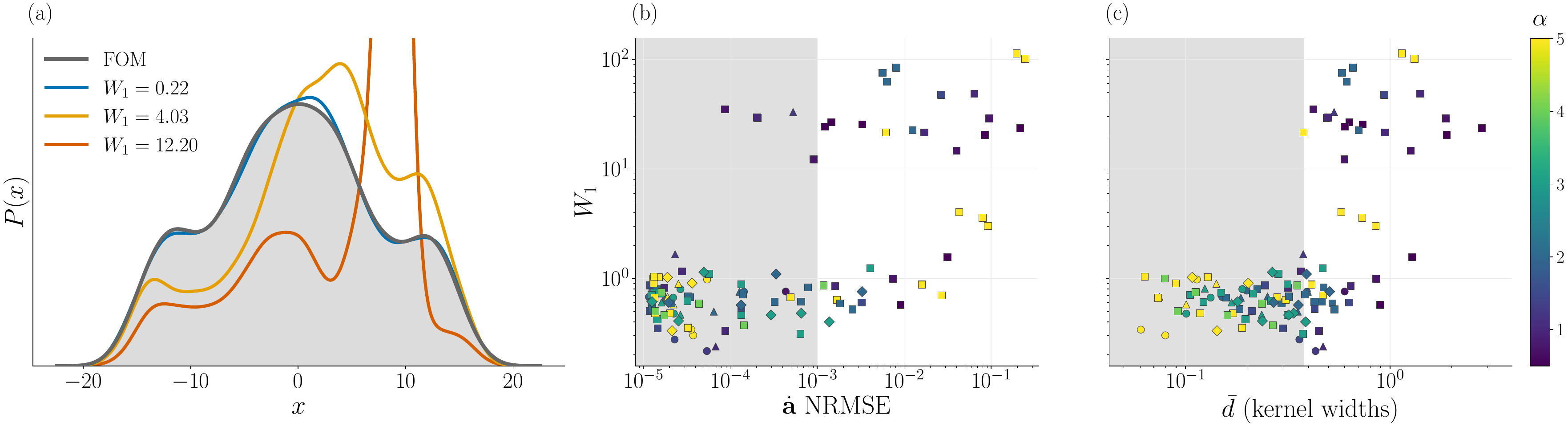}
	\caption{%
		Long-term fidelity against two a-priori measures on Lorenz-63.
		\emph{(a)} density $P(x)$ of three models across the range of $W_1$, against the full-order reference (shaded). 
		\emph{(b)} $W_1$ against the test NRMSE of $\dot{\mathbf a}$.
		\emph{(c)} $W_1$ against the coverage $\bar{d}$. 
		Colour gives the width $\alpha$ and the marker the number of clusters, $K = 1$ (\mrk[1.5]{\bullet}), $2$ (\mrk{\blacktriangle}), $3$ (\mrk[0.85]{\blacksquare}), $4$ (\mrk{\blacklozenge}).
	}
	\label{fig:l63recovery}
\end{figure}

\Cref{fig:l63recovery}(a) shows what the score means: at $W_1 = 0.22$ the density of $x$ lies on the
reference, at $4.0$ the support is right but the mass is misplaced, and at $12$ the trajectory has
collapsed onto one lobe.
We count $W_1 > 5$ as a collapse and $W_1 < 1$ as a recovered invariant.

The one-step error is related to this outcome but does not decide it (\Cref{fig:l63recovery}b).
The poor fits mostly collapse, but so do some of the most accurate ones, and the accuracy of the fit
therefore does not rank the models by the fidelity of their statistics.
A better predictor of collapse is the coverage $\bar{d}$, the mean Mahalanobis distance from a
test state to its ten nearest centres, each measured in the metric of its own kernel,
\begin{equation}
	\begin{gathered}
		\bar{d} = \Big\langle \frac{1}{10} \sum_{i \in \mathcal{N}_{10}(\mathbf{a})} d_i(\mathbf{a}) \Big\rangle_{\mathrm{test}},
		\\
		d_i(\mathbf{a}) = \sqrt{(\mathbf{a}-\mathbf{c}_i)^{\top}\,\mathbf{W}_{k(i)}\,(\mathbf{a}-\mathbf{c}_i)},
	\end{gathered}
	\label{eq:dbar}
\end{equation}
where $\mathcal{N}_{10}(\mathbf{a})$ holds the ten centres nearest to $\mathbf{a}$ under $d_i$ and the
average runs over the test states.
The distance $d_i$ is the one of \Cref{eq:aniso-rbf}, so $\bar{d} = 1$ is one kernel width. No 
model with $\bar{d} < 0.38$ collapses (\Cref{fig:l63recovery}c), so the accurate models that do
must be the ones whose narrow kernels leave voids in the reduced space.
The bound is sufficient but not necessary: many models beyond it still recover the measure.

Recovering the invariant measure is the primary requirement, but the marginal densities do not test
the rates at which the flow stretches and contracts \citep{pathak2017}.
A field that is to be used to study the dynamics, as in the stability analysis and control design of
\Cref{sec:intro}, has to reproduce these rates as well \citep{margazoglou2023}.
The standard measure of these rates is the spectrum of Lyapunov exponents, the mean exponential rates
at which infinitesimal perturbations grow or decay along each direction of the state space
\citep{eckmann1985, benettin1980}.

For Lorenz-63 the spectrum is well known, $\Lambda_1 = 0.906$, $\Lambda_2 = 0$ and
$\Lambda_3 = -14.57$ \citep{sprott2003}, and each exponent reflects one property of the attractor
\citep{eckmann1985}.
The positive exponent is the divergence of neighbouring trajectories that makes the system chaotic,
the zero exponent belongs to the direction of the flow, and the negative one is the contraction onto
the attractor, which makes it thin.
The field is explicit, so its Jacobian is available in closed form,
\begin{equation}
	\mathbf{J}(\mathbf{a}) = \frac{\partial \mathbf{f}}{\partial \mathbf{a}}
	= -\sum_{i=1}^{n_c} \phi_i(\mathbf{a}) \, \boldsymbol{\xi}_i \,
	(\mathbf{a} - \mathbf{c}_i)^{\top} \mathbf{W}_{k(i)},
	\label{eq:jacobian}
\end{equation}
and the exponents follow from the method of \citet{benettin1980}, applied along a trajectory of
the model without the corrector over about $2000$ time units.
The exponents depend on the coverage as well: among the models that recover the measure, $W_1 < 1$,
the median error in $\Lambda_1$ is $0.5\%$ and $\Lambda_2$ stays within $0.004$ of zero.

\begin{figure}[pos=H]
	\centering
	\includegraphics[width=\textwidth]{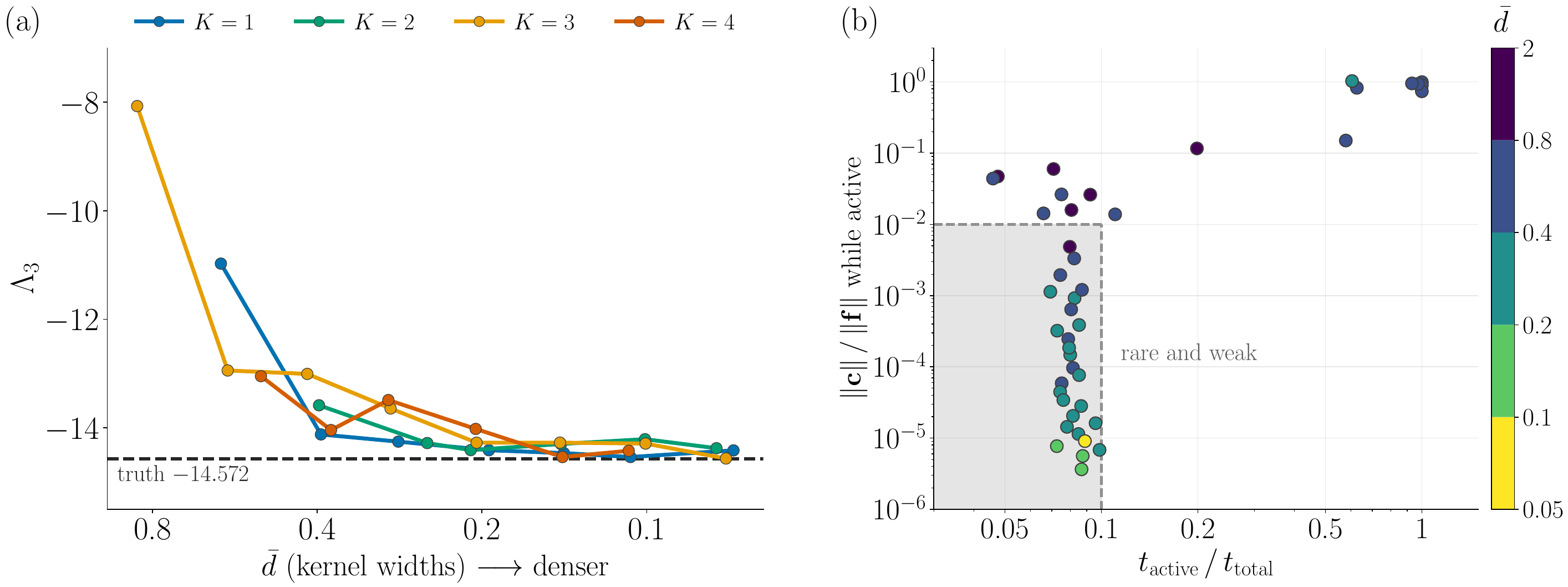}
	\caption{%
		\emph{(a)} contracting Lyapunov exponent $\Lambda_3$ against the coverage $\bar{d}$.
		 Each line is the mean of $\Lambda_3$ over the models of one $K$ in each bin of $\bar{d}$.
		 The dashed line is the reference $\Lambda_3 = -14.572$. 
		 \emph{(b)} duty cycle of the corrector over the $500$ time 
		 units of each free run: the fraction $t_{\mathrm{active}} / t_{\mathrm{total}}$ on which it acts, against the mean of 
		 $\lVert \mathbf{c} \rVert / \lVert \mathbf{f} \rVert$ over that fraction, coloured by $\bar{d}$. 
	}
	\label{fig:l63lyap}
\end{figure}

The contracting exponent is more demanding (\Cref{fig:l63lyap}a).
At the coverage that suffices for the measure, $\bar{d} \approx 0.4$, the mean error is still about
$10\%$, always on the side of too little contraction.
The error falls steadily as the kernels grow denser and settles within $2\%$ for $\bar{d} < 0.1$.
The distance between a state and its nearest kernels therefore sets both the invariant measure
and the stability of the recovered field.

The exponents are computed along a trajectory of the bare field, \Cref{eq:rbf-model}, so they judge
the regression alone, with no contribution from the corrector.
The corrector has little part in the free runs either (\Cref{fig:l63lyap}b).
Every model that recovers $\Lambda_3$ to better than $10\%$ falls in the marked corner of the
panel: the corrector is active on some $8\%$ of the run, and while active it adds about $10^{-5}$
of the field.

A field whose kernels cover its own dynamics therefore leaves the corrector little to do, and the
correction is then a safeguard against the rare excursion rather than a part of the model.
\subsection{Lorenz-96}
\label{sec:results:l96}
Recurrent networks and reservoir computers are the standard non-intrusive forecasters
of chaotic systems, and they often forecast further ahead than an explicit model can.
The case for an explicit field then rests on whether it stays close to their skill
while returning a vector field on the reduced coordinates.
This section measures that gap against the published results of \citet{vlachas2020}.

The comparison runs on the Lorenz-96 benchmark \citep{lorenz1996, lorenz1998}, which describes
a scalar atmospheric variable on a periodic latitude circle,
\begin{equation}
	\frac{\mathrm{d} x_j}{\mathrm{d} t}
	= (x_{j+1} - x_{j-2})\,x_{j-1} - x_j + F,
	\qquad j = 0, \dots, J-1,
	\label{eq:lorenz96}
\end{equation}

with periodic indices $x_{-1} = x_{J-1}$ and $x_{-2} = x_{J-2}$.
The parameters are $J = 40$ and the two forcing regimes $F = 8$ and $F = 10$, both chaotic,
with maximal Lyapunov exponents $\Lambda_1 \approx 1.68$ and
$\Lambda_1 \approx 2.27$ respectively \citep{vlachas2020}.

The system is integrated with a step $h = 0.01$, as in \citet{vlachas2020}, and sampled
every $\Delta t = 0.1$ time units, over $2000$ time units after a transient of $1000$.
It is single-scale, with no spectral gap, so the variance discarded by truncation
is spread thinly over many modes rather than held in a fast subsystem, and the reduced
observable is the harder target.
The models are tested on the full state, $r = 40$, and on a reduced observable,
$r = 35$ with $\varepsilon \approx 0.96$, formed by the singular-value decomposition
of \Cref{sec:method:coords}, on a record of the same length as that of \citet{vlachas2020}.
\Cref{app:extra} reports the convergence of the radial-basis fit and the number of centres
on both observables.

The chaotic states of \Cref{eq:lorenz96} keep the spatial scale of the first instability,
wavenumber eight at $J = 40$, in a disordered train drifting towards decreasing site index
\citep{vankekem2018, kerin2022}.
One cycle of its phase takes $1.77$ time units at $F = 8$ and $1.64$ at $F = 10$.

Two properties are measured.
Short-term skill is the following of a trajectory, through the normalised root-mean-square
error and the valid prediction time defined below.
Long-term skill is the invariant measure, through the probability density of the energy
and the power spectral density of the reduced state.
Both are measured on the full state and on the reduced observable, to show how much of each
survives the truncation.

\begin{figure}[pos=H]
	\centering
	\includegraphics[width=.8\textwidth]{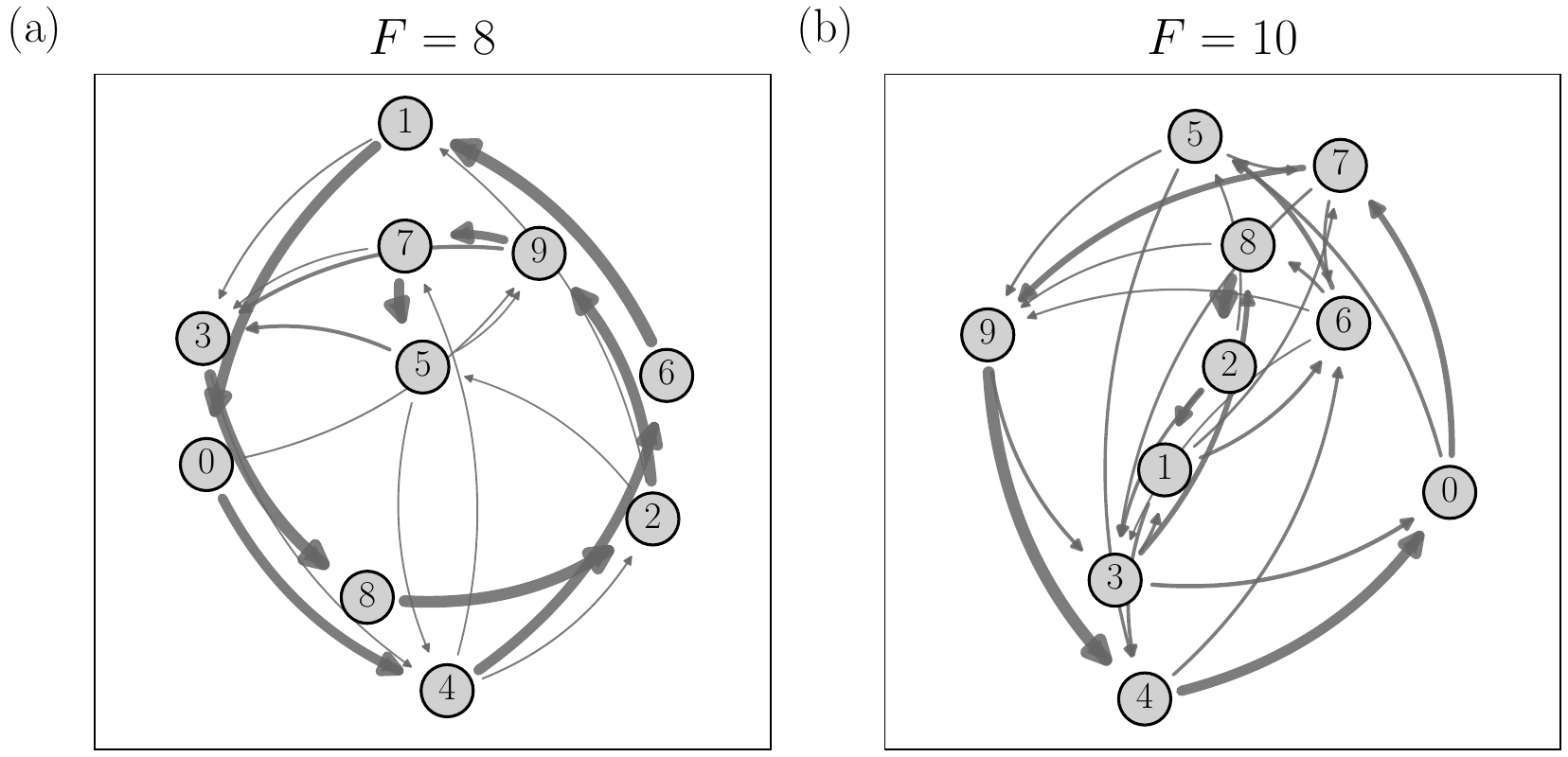}
	\caption{%
		Transition structure of the Lorenz-96 partition at \emph{(a)} $F = 8$ and \emph{(b)} $F = 10$,
		laid out in the phase of the travelling wave.
		Arrows are the transitions, thickening with their frequency. }
	\label{fig:app:lor:energy}
\end{figure}
The partition finds that wave (\Cref{fig:app:lor:energy}).
The centroids are cyclic shifts of one another, so each cluster holds one phase of the train, and
the arrows follow its drift.
The wavenumber and the period are unchanged, but two things blur the ordering at $F = 10$: the
chaos is faster, $\Lambda_1$ rising from $1.68$ to $2.27$, and the wave holds less of the state,
mode eight falling from $11.9$ to $9.9\%$ of the power.
\begin{figure}[pos=H]
	\centering
	\includegraphics[width=\textwidth]{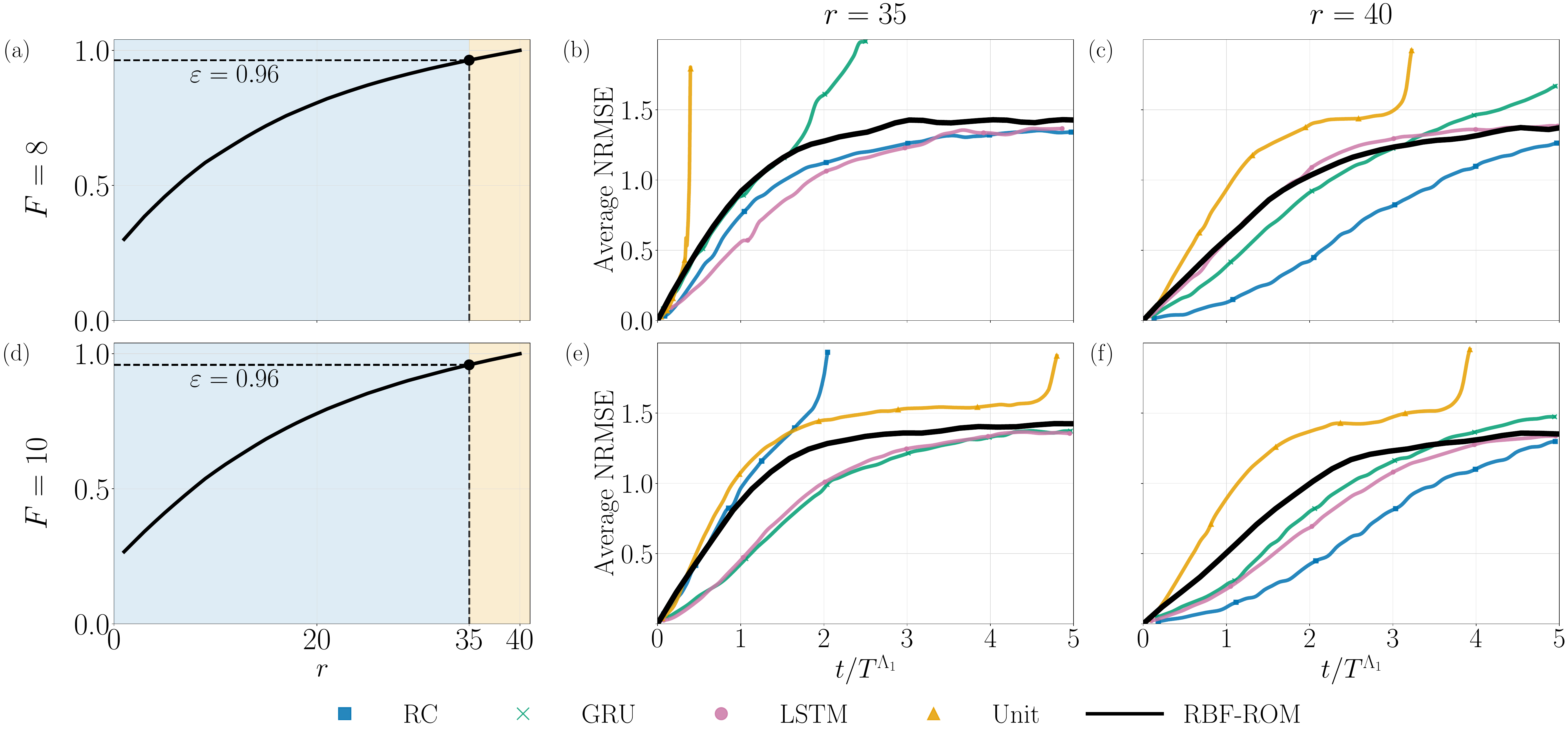}
	\caption{%
		Lorenz-96 forecasting against the data-driven baselines of \citet{vlachas2020}, for $F = 8$ (top) and $F = 10$ (bottom). 
		\emph{(a, d)} cumulative variance against the number of singular-value modes.
		\emph{(b, e)} mean forecast error \Cref{eq:nrmse} on the reduced observable and \emph{(c, f)} on the full state, 
		against $t / T^{\Lambda_1}$. The radial-basis model (black) over the reservoir computer (RC), gated recurrent unit (GRU), 
		long short-term memory (LSTM) and unitary (Unit) networks, each at its best configuration.
	}
	\label{fig:l96forecast}
\end{figure}
The forecast $\hat{\mathbf{a}}(t)$ is compared against the truth $\mathbf{a}(t)$
through the normalised root-mean-square error
\begin{equation}
	\mathrm{NRMSE}(t)
	= \sqrt{\frac{1}{r} \sum_{i=1}^{r}
		\frac{\bigl(\hat{a}_i(t) - a_i(t)\bigr)^2}{s_i^2}},
	\label{eq:nrmse}
\end{equation}
where $s_i$ is the standard deviation of coordinate $a_i$ over the attractor,
so that each coordinate contributes on its own scale.
The valid prediction time \citep{vlachas2020} is then the first horizon at which this error crosses a
tolerance $\delta$, measured in Lyapunov times,
\begin{equation}
	\mathrm{VPT}
	= \frac{1}{T^{\Lambda_1}}
		\max\bigl\{\, \tau : \mathrm{NRMSE}(t) < \delta
		\ \text{for all}\ t \le \tau \,\bigr\},
	\label{eq:vpt}
\end{equation}
with $\delta = 0.5$ and $T^{\Lambda_1} = 1/\Lambda_1$ the Lyapunov time, averaged
over $100$ initial conditions drawn from the attractor.

Against the recurrent and reservoir baselines at their best configurations, as reported by
\citet{vlachas2020}, the error of the radial-basis field grows within the band the four of them
span, over the whole horizon (\Cref{fig:l96forecast}b, c, e, f).

\Cref{tab:l96vpt} compares the valid prediction time of each architecture, at the
median and at the best of its hyperparameter grid \citep{vlachas2020}, with that of
the radial-basis model.
The single fit matches or exceeds every median configuration but one, the LSTM at
$F = 10$ on the reduced observable, which leads it by $0.01$.
It exceeds the best unitary network in two of the four cases and ties it in a third.
The best reservoir computer, LSTM and GRU lead it in every case: the reservoir computer by $27$ to
$33\%$ on the reduced observable and by $2.5$ to $2.8$ times on the full state, the LSTM by $1.7$
to $2.6$ times and by $18$ and $85\%$, and the GRU by $2.2$ to $2.7$ times and by $63$ and
$70\%$.
A network at a typical configuration therefore forecasts as far as the radial-basis
field, and a tuned one further.
The field is the simpler model: one regression with two scalar knobs, the width and
the budget, in place of an architecture search, and it returns an explicit vector
field on the reduced coordinates.
\begin{figure}[pos=H]
	\centering
	\includegraphics[width=\textwidth]{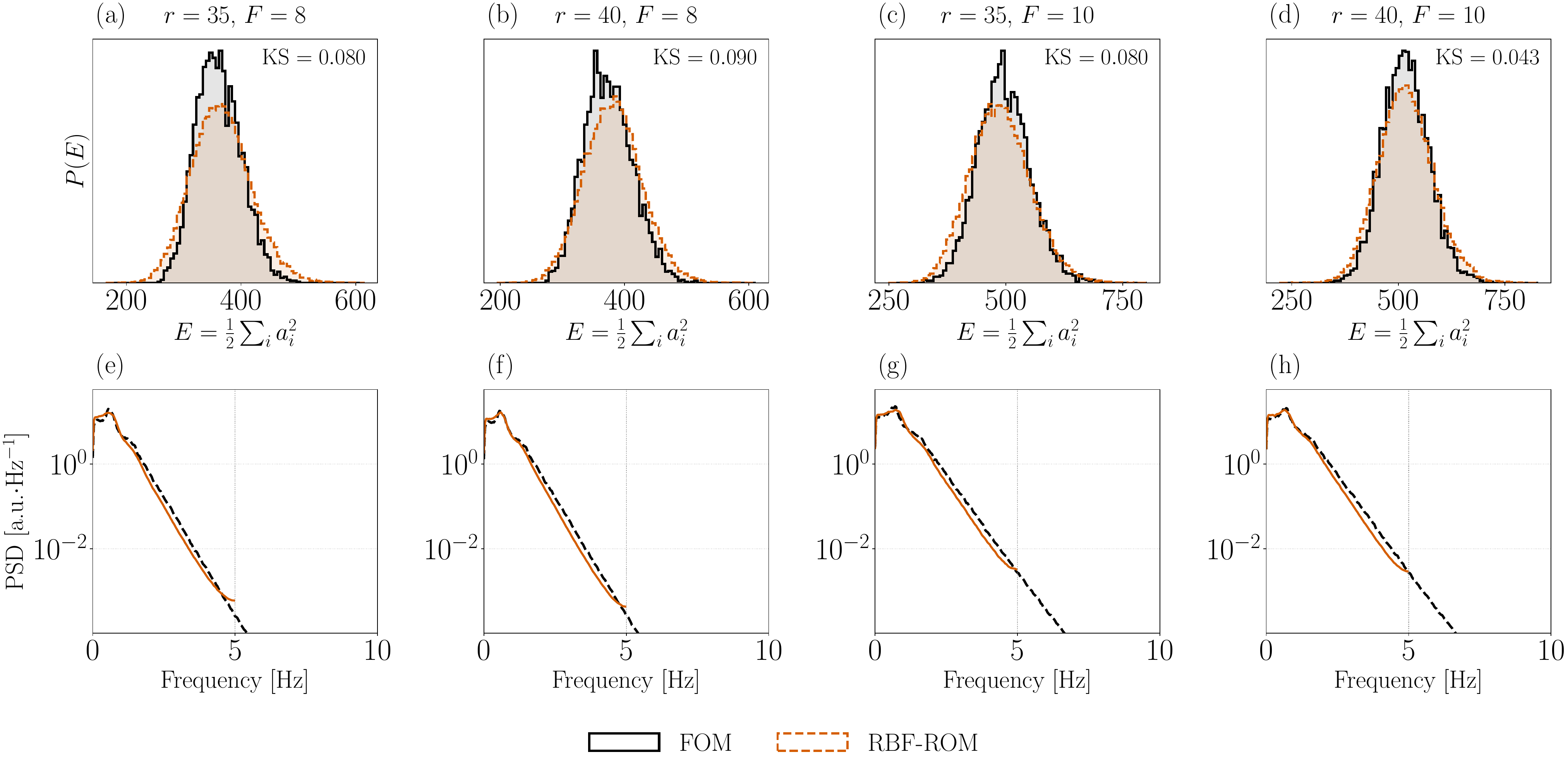}
	\caption{%
		Lorenz-96 invariant measure of the reduced coordinates, full-order system (FOM) against the radial-basis model, 
		for $r = 35$ and $r = 40$ at both forcings. \emph{(a--d)} energy density $E = \tfrac{1}{2}\sum_i a_i^2$, 
		with the Kolmogorov--Smirnov distance annotated. \emph{(e--h)} power spectral density of the reduced state; 
		the dotted line marks the sampling Nyquist frequency.
	}
	\label{fig:l96invariant}
\end{figure}

\begin{table}[pos=H]
	\caption{%
		Valid prediction time \Cref{eq:vpt} on Lorenz-96, in Lyapunov times. 
		Baselines are the median of the hyperparameter grid of each architecture in \citet{vlachas2020}, 
		with its best configuration in brackets; the radial-basis model is a single fit, 
		given as the mean over $100$ initial conditions plus or minus one standard deviation. 
		Green marks the baseline values that the radial-basis mean of the row exceeds.
	}
	\label{tab:l96vpt}
	\footnotesize
	\setlength{\tabcolsep}{3pt}
	\begin{tabular*}{\tblwidth}{@{\extracolsep{\fill}}ccccccc@{}}
		\toprule
		$F$ & $r$ & RC & GRU & LSTM & Unit & Radial basis \\
		\midrule
		$8$  & $35$ & \win{$0.32$} $(0.56)$ & \win{$0.37$} $(0.98)$ & \win{$0.37$} $(0.75)$ & \win{$0.34$} $(0.44)$ & $0.44 \pm 0.14$ \\
		$8$  & $40$ & \win{$0.80$} $(2.32)$ & \win{$0.56$} $(1.35)$ & \win{$0.44$} $(0.98)$ & \win{$0.34$} \win{$(0.58)$} & $0.83 \pm 0.25$ \\
		$10$ & $35$ & \win{$0.35$} $(0.61)$ & \win{$0.43$} $(1.23)$ & $0.47$ $(1.18)$ & \win{$0.39$} $(0.50)$ & $0.46 \pm 0.16$ \\
		$10$ & $40$ & \win{$0.83$} $(2.36)$ & \win{$0.71$} $(1.60)$ & \win{$0.66$} $(1.74)$ & \win{$0.41$} \win{$(0.63)$} & $0.94 \pm 0.27$ \\
		\bottomrule
	\end{tabular*}
\end{table}

The long-term statistics are compared in \Cref{fig:l96invariant} through the
probability density of the energy $E = \tfrac{1}{2}\sum_i a_i^2$ and the power
spectral density of the reduced state.
The radial-basis model reproduces the energy density at both forcings and both
observables, with a Kolmogorov--Smirnov distance between $0.04$ and $0.09$, and its
spectra follow the full-order system up to the Nyquist frequency of the sampling.

We do not claim this as a point of superiority over the data-driven baselines,
which, properly built, also reproduce the Lorenz-96 climate \citep{vlachas2020}.
The model matches the invariant measure while remaining a single global, explicit and
differentiable field, fitted non-intrusively by one regression.

On Lorenz-96, then, the model forecasts as far as these architectures at their typical
configurations and trails their best, it reproduces the long-term statistics, and it does so as
an analysable field rather than an opaque one.
The partition invites the same reading: built from snapshots alone, and only to place the kernels,
it orders its clusters by the phase of the travelling wave, so the reduced model can be
interrogated as well as integrated.
\subsection{Kuramoto--Sivashinsky and Kolmogorov flow}
\label{sec:results:pde}

\Cref{sec:results:l63,sec:results:l96} assess the method on ordinary differential
equations.
The two systems that follow are partial differential equations, whose governing equations are
available, and they allow the comparison with the intrusive alternative: the Galerkin
reduced-order model obtained by projecting the equations onto the POD basis, in its
global (g-ROM) and quantised-local (ql-ROM) forms \citep{colanera2025}.
In both cases the model is run freely over the whole test record, started from its
first state.

\begin{figure}[pos=H]
	\centering
	\includegraphics[width=.9\textwidth]{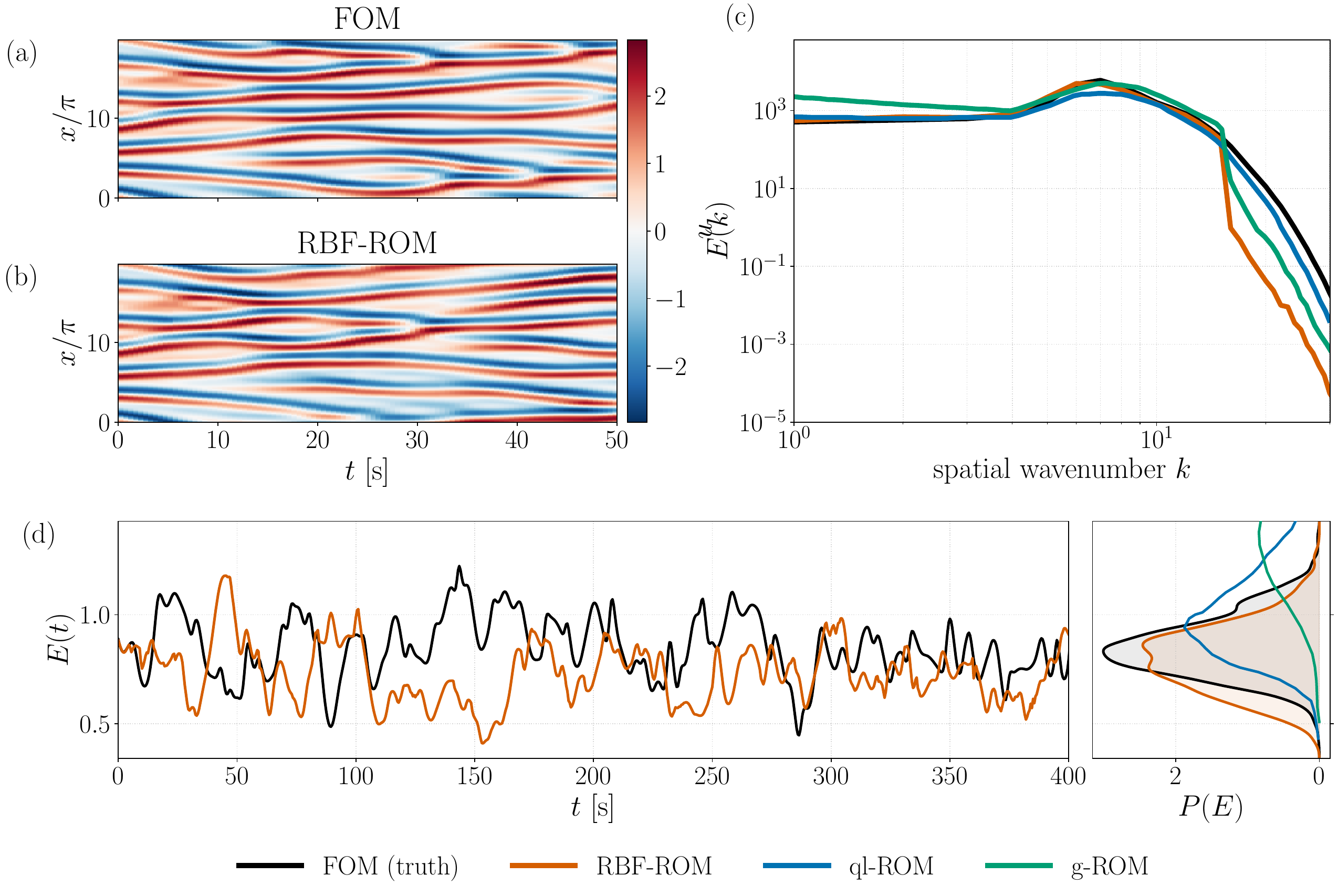}
	\caption{%
		Kuramoto--Sivashinsky in the chaotic regime ($L = 20\pi$, $\nu = 1$). \emph{(a,~b)}~space-time evolution of the 
		full-order solution and a free run of the model. \emph{(c)}~spatial energy spectrum $E(k)$. \emph{(d)}~reduced energy 
		$E(t)$ for the full-order system and the model, with the density $P(E)$ of all four models at right. 
		The ql-ROM and g-ROM are the intrusive quantised-local and global Galerkin models of \citet{colanera2025}. 
	}
	\label{fig:ks}
\end{figure}

The first is the Kuramoto--Sivashinsky equation \citep{kuramoto1976, sivashinsky1977},
\begin{equation}
	\frac{\partial u}{\partial t}
	+ u \frac{\partial u}{\partial x}
	+ \frac{\partial^2 u}{\partial x^2}
	+ \nu \frac{\partial^4 u}{\partial x^4} = 0,
	\qquad x \in (0, L],
	\label{eq:ks}
\end{equation}
with periodic boundary conditions, in its chaotic regime, $L = 20\pi$ and $\nu = 1$, a
broadband and statistically stationary state \citep{colanera2025}.
The equation is solved on $128$ grid points, integrated with a step $h = 0.05$ and sampled
every $\Delta t = 0.5$ time units, over $30\,000$ time units after a transient of $1500$.
\begin{figure}[pos=H]
	\centering
	\includegraphics[width=.8\textwidth]{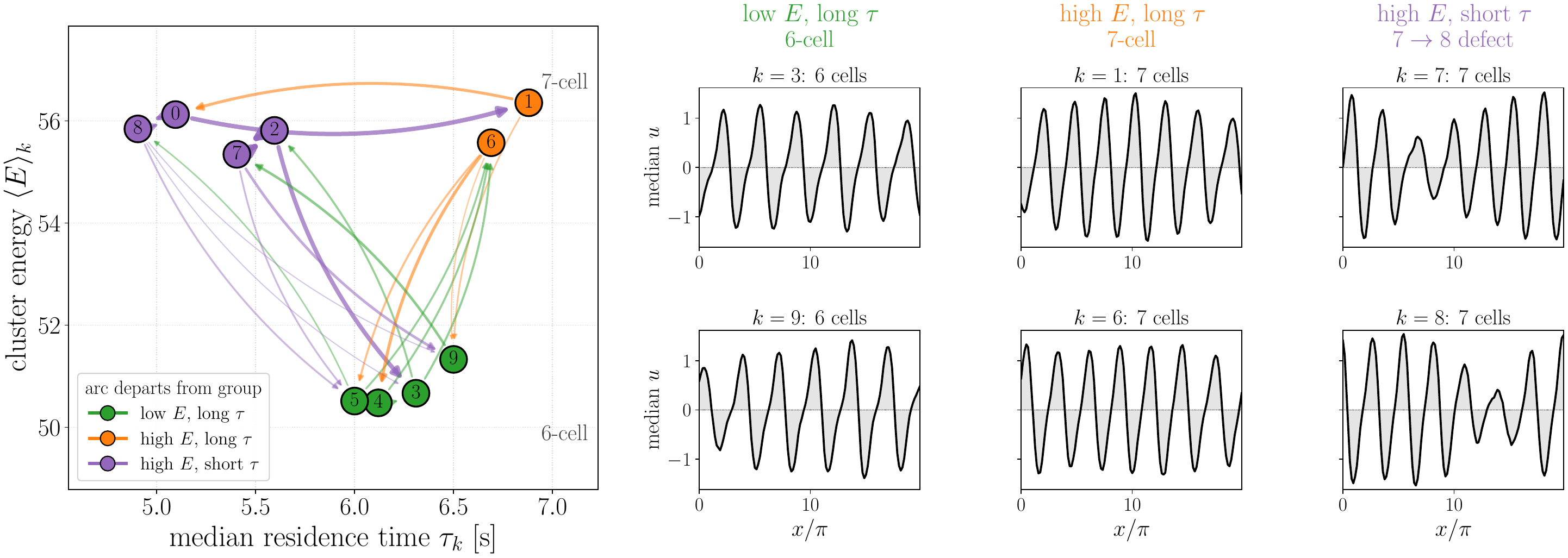}
	\caption{%
		Energetic structure of the Kuramoto--Sivashinsky partition. 
		\emph{Left:} cluster mean energy $\langle E \rangle_k$ against median residence time $\tau_k$, 
		in three groups. Arrows show the dominant transitions. 
		\emph{Right:} median profiles of representative clusters, showing the six- and seven-cell states and the local defect.
	}
	\label{fig:app:ks:energy}
\end{figure}
The model retains $r = 30$ reduced coordinates, which carry $99\%$ of the variance, and the
partition is set at $K = 10$ (\Cref{app:extra}).
As everywhere in this work the weights are fitted over the whole library in one solve, so the field
itself is a single global one.
Set beside a space-time portrait of the full-order solution, a free run sustains the cellular chaos
characteristic of the equation, with the same spatial scale and the same irregular merging and
splitting of structures, throughout the run (\Cref{fig:ks}a,~b).

The states of \Cref{eq:ks} are cellular, with a mean wavelength close to that of the fastest
instability of the flat state, $2\pi\sqrt{2}$, which fits $7.07$ times into $L = 20\pi$
\citep{hyman1986, cvitanovic2010}.
The dominant wavenumber of the record runs from $k = 5$ to $9$, with $64\%$ of it on six or seven
cells, which merge and split into one another.

These configurations are better separated by splitting the clusters by energy and by the median
time spent in them, $\tau_k$.
The six- and seven-cell configurations then fall at low and high energy respectively
(\Cref{fig:app:ks:energy}): the cluster means are $50.7$ and $55.8$, each family spread by about one unit against
a gap of five between them.
A third group sits at the seven-cell energy but is held for five time units against six and a half,
and carries twice the eighth-cell content of the rest.
Within a family the clusters tile the phase of the pattern almost perfectly, reaching $100$ and
$91\%$ of the concentration that evenly spaced bins would give, against $64\%$ on Lorenz-96.

The spatial energy spectrum, \Cref{fig:ks}(c), is compared over the resolved band.
The $30$ POD modes resolve the wavenumbers up to $k = 15$ (\Cref{app:extra}), so
the reconstruction carries no energy beyond it by construction, and the test is the
agreement below it: there the model follows the full-order spectrum through the
energetic wavenumbers around the peak at $k \approx 7$.
The ql-ROM, with the same $r = 30$, extends further into the tail because its bases are local, and
the centroid of each cluster carries content beyond the global cut.
The g-ROM holds excess energy at the largest scales: the truncation removes the small-scale sink,
so energy piles up at low $k$, the same defect that over-disperses its energy
density.
The probability density of the field
energy, \Cref{fig:ks}(d), is the more telling diagnostic: the density of the model
sits on that of the full-order system, as close as the intrusive ql-ROM, while the
global projection over-disperses the energy, spreading the density well beyond its
true support.
The non-intrusive field thus tracks the invariant measure of the energy as closely as the
equation-based local model, without the over-dispersion of the global one.

The second develops in two space dimensions,
the Kolmogorov flow: the two-dimensional Navier--Stokes equations on a periodic
square, driven by a stationary body force along $x$ \citep{colanera2025},
\begin{equation}
	\begin{aligned}
		\nabla \cdot \mathbf{u} &= 0,
		\qquad \mathbf{x} \in [0, 2\pi)^2, \\
		\frac{\partial \mathbf{u}}{\partial t}
		+ (\mathbf{u} \cdot \nabla)\mathbf{u}
		&= -\nabla p + \frac{1}{Re}\nabla^2 \mathbf{u} + \mathbf{g},
	\end{aligned}
	\label{eq:kol}
\end{equation}
with periodic boundary conditions, where $\mathbf{u} = (u, v)$ is the velocity, $p$ the
pressure, $Re$ the Reynolds number and $\mathbf{g} = (\sin 4y, 0)$ the forcing.
The resulting vector field is an incompressible flow which, at the Reynolds number
considered, is quasiperiodic, so the set to reproduce is a torus rather than a
broadband attractor.
The force sustains a laminar shear flow that loses stability as the Reynolds number
grows, through a sequence of bifurcations \citep{platt1991} that ends in a turbulent
state \citep{chandler2013}.
The Reynolds number here is $Re = 20$, early in this sequence.

The flow is solved with the pseudospectral solver KolSol \citep{kelshaw2022kolsol} on a
$64 \times 64$ grid, integrated with a step $h = 0.01$ and sampled every $\Delta t = 0.5$
time units, over $10^4$ time units.
The model retains $r = 23$ coordinates, again $99\%$ of the variance (\Cref{app:extra}).

\begin{figure}[pos=H]
	\centering
	\includegraphics[width=\textwidth]{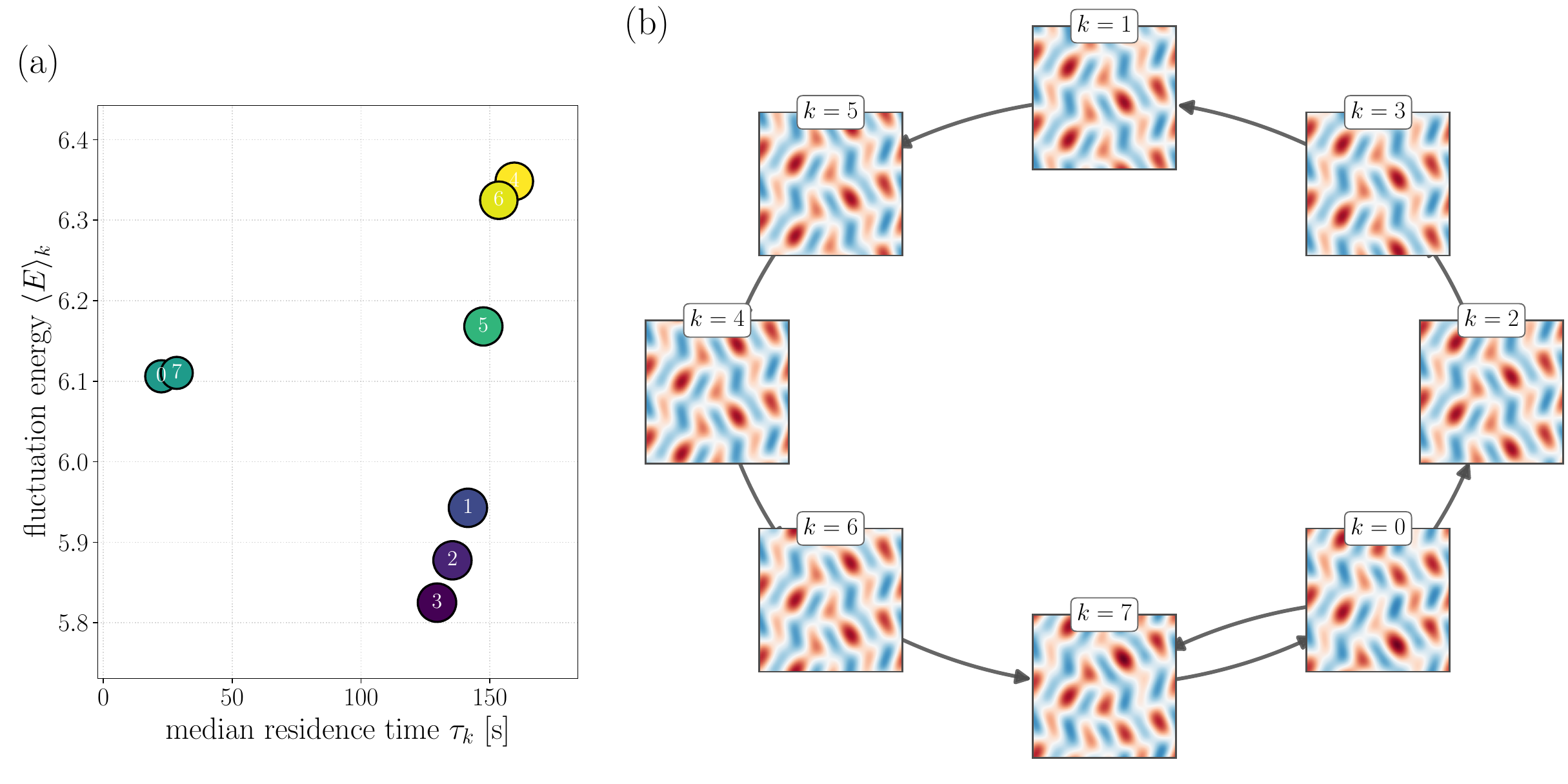}
	\caption{%
		Structure of the Kolmogorov partition. \emph{(a)} cluster fluctuation energy $\langle E \rangle_k$ against median residence time $\tau_k$; six clusters are held for some $145$ time units, two for $25$. \emph{(b)} cluster-mean vorticity fields in a ring, with arrows giving the dominant transitions.
	}
	\label{fig:app:kol:energy}
\end{figure}

At $Re = 20$ the flow is quasiperiodic, and its two leading modes are a degenerate pair carrying
$39$ and $38\%$ of the variance: the torus is dominated by one circle, whose phase turns once in
about a thousand time units.

The partition tiles that circle.
Laid out in the order the clusters are visited, the cluster-mean vorticity fields resolve the phase
to $99\%$ of what evenly spaced bins would give, six of the eight clusters stepping one bin per
transition (\Cref{fig:app:kol:energy}).
The remaining two are one region cut in two: their centroids lie two and a half times closer than
any other pair, and they exchange with each other in three switches of four, which is what their
short residence records.

\begin{figure}[pos=H]
	\centering
	\includegraphics[width=\textwidth]{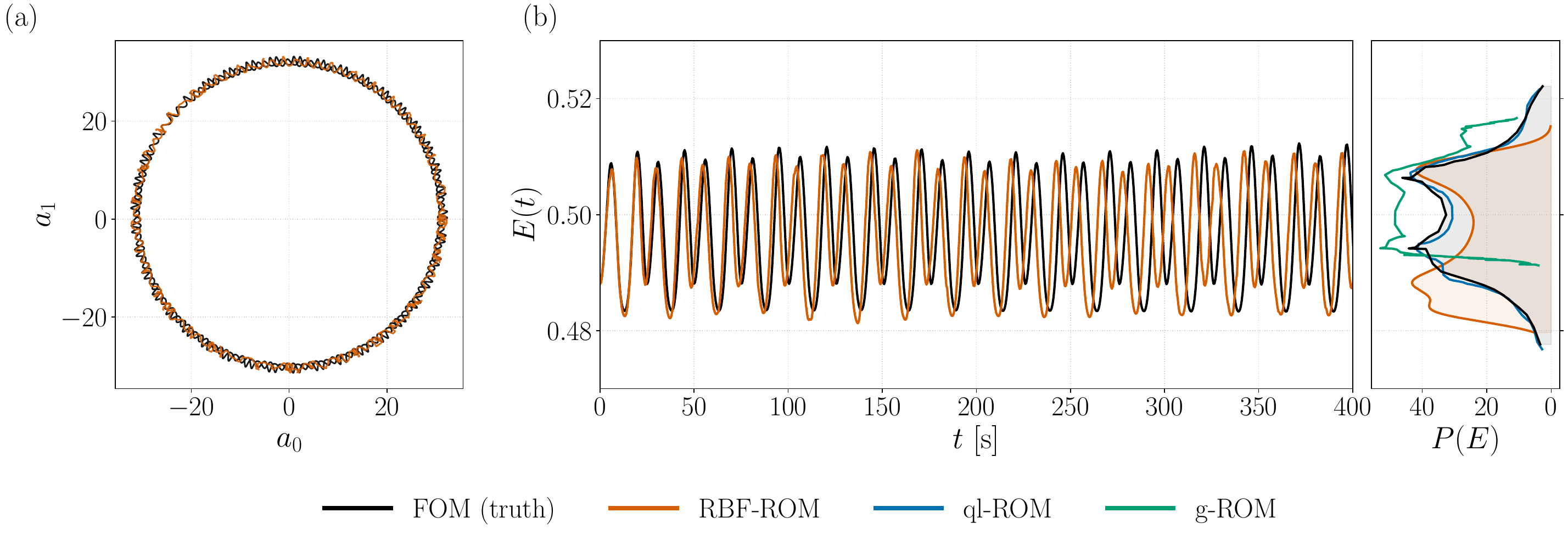}
	\caption{%
		Kolmogorov flow at $Re = 20$ in the quasiperiodic regime. \emph{(a)}~projection onto the two leading reduced coordinates,
		 full-order (black) and model (red), tracing the invariant torus. 
		 \emph{(b)}~reduced kinetic energy $E(t)$ and its density $P(E)$. 
		 The ql-ROM and g-ROM are the Galerkin models of \citet{colanera2025}.
	}
	\label{fig:kol20}
\end{figure}

Projected onto its two leading coordinates $a_0, a_1$, the model traces the same closed torus as
the full-order flow, neither spiralling inward to a fixed point nor drifting outward
(\Cref{fig:kol20}a).

The probability density of the kinetic energy, \Cref{fig:kol20}(b), agrees with the full-order
statistics and with the ql-ROM.
The density is narrow, as a quasiperiodic measure must be, and the model recovers its position and
width.
The full-order data here come from a direct numerical simulation whose energy density is modestly
wider than that reported by \citet{colanera2025}, and the Galerkin densities are accordingly
rescaled to a common variance, so that the comparison is one of shape rather than of absolute
spread.

Across a chaotic and a quasiperiodic flow, then, the radial basis field reproduces
the energy density as accurately as the intrusive quantised-local Galerkin model,
and on Kuramoto--Sivashinsky the energy spectrum over the resolved band as well, and
it does so without ever invoking the governing equations.

\section{Conclusions and outlook}
\label{sec:conclusions}
The aim of this work was to learn the dynamics of a chaotic system from data alone, with no knowledge
of the governing equations and no assumed analytic form, and to return that dynamics as a vector
field that can be integrated, differentiated and examined.
To this end the attractor is partitioned by clustering, the principal directions of each cluster
shape an anisotropic radial basis, and the weights of the whole library are fitted in one linear
solve.
The shape of the field is local and its dynamics global: a state crosses from one region to the next
without any switch of model, and the extent of the kernels is the only thing the partition controls.
The kernels decay away from their centres, so the field also measures its own distance from the data,
and a kinematic corrector uses that distance to hold a long integration on the attractor.

The accuracy of the fit does not decide whether the integrated field reproduces the system.
On Lorenz-63, where the attractor and the Lyapunov spectrum are known in advance, the one-step error
is set almost entirely by the condition number of the library, and some of the most accurate fits
are among those whose free runs collapse onto one lobe.
What decides the outcome is the coverage $\bar{d}$, the mean distance from a state to its nearest
kernels in the metric of those kernels: no model that held it below $0.38$ kernel widths collapsed,
and the contracting exponent settled within $2\%$ of its reference value once it fell below $0.1$.
The number of clusters acts through that distance rather than on its own, since partitioning lowers
the condition number at equal width and budget and so allows a denser cover.

Against other data-driven and intrusive models the field is competitive, without being the most
accurate forecaster.
On Lorenz-96 the valid prediction time matched or exceeded a typical configuration of every recurrent
and reservoir architecture tested by \citet{vlachas2020}, while the best-tuned configurations of
three of the four forecast further, and the energy density and the power spectra were reproduced on
both the full state and the reduced observable.
On the chaotic Kuramoto--Sivashinsky equation and the quasiperiodic Kolmogorov flow, and without any
use of the governing equations, the model reproduced the energy distribution as closely as the
intrusive quantised-local Galerkin model \citep{colanera2025}, and more closely than a global
Galerkin projection of the same reduced dimension, which over-disperses the energy.
On Kuramoto--Sivashinsky the resolved spectrum agrees as closely as well.

Partitioning an attractor and fitting within each region is shared with the cluster-based
reduced-order models of \citet{kaiser2014} and \citet{colanera2025}, and the regression itself is
the standard scattered-data interpolant \citep{broomhead1988, micchelli1986}.
Specific to this work are three things.
The cluster covariance shapes the kernels while the dynamics remain a single field fitted over one
library, so the state never changes model.
The coverage, and not the error of the fit, is what decides the recovery of the invariant measure.
A field fitted with no use of the governing equations reproduces that measure as closely as an
intrusive local Galerkin model of the same reduced dimension.

These results carry limits of their own.
The model treats the retained coordinates as self-contained, fitting the velocity as a function of
the state alone.
Truncation leaves a memory behind that the field cannot see, and its price shows on Lorenz-96, where
the same construction forecasts about twice as far on the full state, which is closed, as on the
reduced observable, which is not.
How far the method reaches in dimension is a question of budget.
The coverage the fit needs is a fill distance in the sense of scattered-data approximation, and for
centres spread over a set of dimension $D$ the budget that holds it fixed grows exponentially in
$D$ \citep{schaback1995, fasshauer2007}.
Two things lower that exponent here.
The centres sit on the data, so $D$ is the dimension of the attractor and not that of the reduced
space, and each kernel is elongated along the attractor, so it covers a sheet of it rather than a
ball.
Both are what make forty coordinates tractable at all, and neither is measured, so the dimension at
which the budget becomes prohibitive is unknown.

The partition is the least adaptive part of the construction.
$K$-means cuts the reduced coordinates into Voronoi cells, which sit awkwardly on a thin and curved
attractor.
The criterion for the number of regions had to be set by convention on Lorenz-96, and the local
metric degrades as a cluster flattens and its covariance loses rank.
The corrector, finally, is built from the geometry of the training data and acts during the long
integrations, so the statistics reported are those of the field together with its safeguard.

The construction leaves two directions open.
The field is explicit and differentiable, so the objects that organise a chaotic attractor can be
sought directly in the reduced space: its fixed points by Newton iteration on the right-hand side,
its unstable periodic orbits on the same few coordinates, and the stability of each from the
Jacobian in closed form, computations that are feasible on a reduced state and intractable on the
full one.
None of these was attempted here.
The second direction addresses the limit noted above.
Replacing the kinematic corrector by a learned closure, one that models the effect of the discarded
coordinates on the retained ones, would return the attracting dynamics to the field itself and
remove the only component that is not learned from data.
Beyond these, the construction acts on a set of coordinates and is indifferent to their provenance,
so the same reduced field supports analyses that call on no governing equations, from the topology
of the sampled set \citep{ghrist2008, carlsson2009} to the directed information between coordinates
\citep{schreiber2000}.

One implication is practical.
The coverage is computed from the fitted centres and the training states alone, before any
integration, so a candidate fit can be screened at the cost of a nearest-neighbour search rather
than of a long run.
The threshold measured here is a bound on Lorenz-63 and not a universal one, so it would have to be
established for each system, but the screening itself needs no free run.

In summary, a chaotic system can be given an explicit and differentiable reduced vector field whose
free running reproduces its invariant measure and its Lyapunov spectrum, provided the kernels cover
the attractor densely enough.
Accuracy on the measured velocities is not a reliable indicator of that.

% Use the style of numbering in square brackets.
% If nothing is used, default style will be taken.
%\begin{enumerate}[a)]
%\item
%\item
%\item
%\end{enumerate}

%\begin{itemize}
%\item
%\item
%\item
%\end{itemize}

%\begin{description}
%\item[]
%\item[]
%\item[]
%\end{description}

%\clearpage %%Remove this from your manuscript

% Figure (template example, kept for reference)
%\begin{figure}%[]
%  \centering
%%    \includegraphics{}
%    \caption{}\label{fig1}
%\end{figure}

% Table (template example, kept for reference)
%\begin{table}%[]
%\caption{}\label{tbl1}
%\begin{tabular*}{\tblwidth}{@{}LL@{}}
%\toprule
%  &  \\ % Table header row
%\midrule
% & \\
% & \\
% & \\
% & \\
%\bottomrule
%\end{tabular*}
%\end{table}

% Uncomment and use as the case may be
%\begin{theorem}
%\end{theorem}

% Uncomment and use as the case may be
%\begin{lemma}
%\end{lemma}


\clearpage

\printcredits

\section*{Data availability}
The data and code that support the findings of this study are openly available at
\url{https://github.com/miketwix373/RBF_ROM}.

\appendix

\section{The trust-region drift corrector}
\label{app:corrector}

This appendix runs the calibration of \Cref{sec:method:integration} on a fitted
model and asks whether the corrector disturbs the dynamics on the attractor that
it is meant to leave alone.
The diagnostics are computed on Lorenz-96 at $F = 8$, on the headline cell
($K = 10$ clusters, $n_c = 4000$ kernels), where the unaided field leaves the
attractor.
The corrector is evaluated inside every Runge--Kutta stage, so the field
integrated is $\mathbf f + \mathbf g$ throughout.

\subsection{Calibration in practice}
\label{app:corrector:calibration}

\begin{figure}[pos=H]
	\centering
	\includegraphics[width=.9\textwidth]{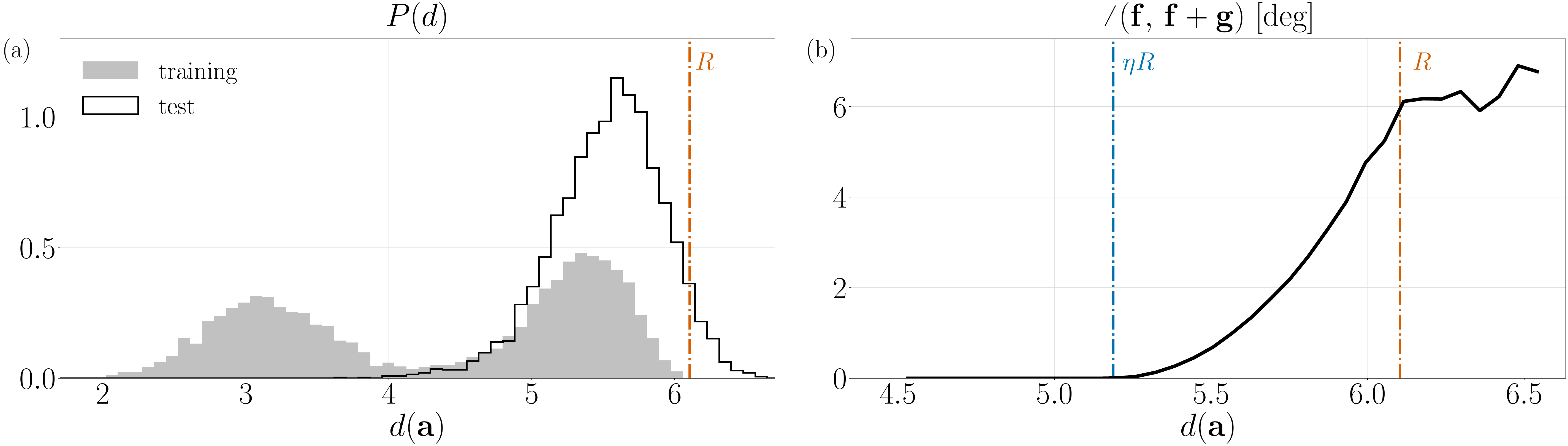}
	\caption{%
		Corrector calibration on Lorenz-96 at $F = 8$, $r = 40$. \emph{(a)} distribution of the nearest-centre distance $d$ over the training and test partitions; the trust radius $R_k$ is the ninety-ninth percentile of $d$ over the snapshots of each cluster. \emph{(b)} mean angle between the learned field $\mathbf f$ and the integrated field $\mathbf f + \mathbf g$ on the test states, as a function of $d$; the corrector engages at $\eta R_k$ and reaches its nominal strength at $R_k$.
	}
	\label{fig:app:calibration}
\end{figure}

\Cref{fig:app:calibration} applies the recipe of \Cref{eq:Rk,eq:kk} to this model.
Panel~(a) sets the trust radius from the distribution of the nearest-centre
distance: the ninety-ninth percentile gives $R_k$ between $6.06$ and $6.15$ across
the ten clusters, so the onset $\eta R_k$ sits at $5.2$.
The clusters agree this closely because Lorenz-96 is statistically homogeneous on
the ring, so every cluster sees the same local geometry.
On an inhomogeneous attractor the radii would separate, which is why the trust region is built per
cluster rather than once for the whole field.
Panel~(b) reports what the correction costs the field: below the onset the
corrector is inactive and the direction of the flow is untouched, and even at the
trust radius it turns the field by about six degrees.

The strength follows from the outward residual in the shell, whose ninety-fifth
percentile is $2.0$ in the whitened coordinates, so \Cref{eq:kk} returns
$\gamma \approx 3$, ranging from $2.5$ to $3.3$ across the clusters.
The residual is about a tenth of the full radial velocity there, whose own
percentile is $19.8$, so cancelling the full velocity would demand
$\gamma \approx 30$, and the worst-case field bound read from \Cref{fig:balance}
is larger still.
That gap is the mark of a healthy fit, and it is what allows the corrector to be
sized from the residual without opposing the dynamics the field already carries.

The runs reported throughout use $\gamma = 10$, some three times the calibrated
value.
\Cref{fig:balance} shows why the setting need only fall in the right range rather
than on a particular value: $\gamma = 1$ already bends the balance inward beyond
$R_k$, while the distortion inside the buffer appears only by $\gamma = 100$.

\subsection{The corrector preserves the invariant measure}
\label{app:corrector:invariant}
The corrector acts inside every long integration, so the statistics reported in the main text are
those of the field together with its safeguard.
This subsection asks whether the safeguard moves them.
A long run of the guarded model is set against the full-order reference, and the full-order state is
then split by whether it lies inside the trust region or in the tail the corrector acts on.

\begin{figure}[pos=H]
	\centering
	\includegraphics[width=\textwidth]{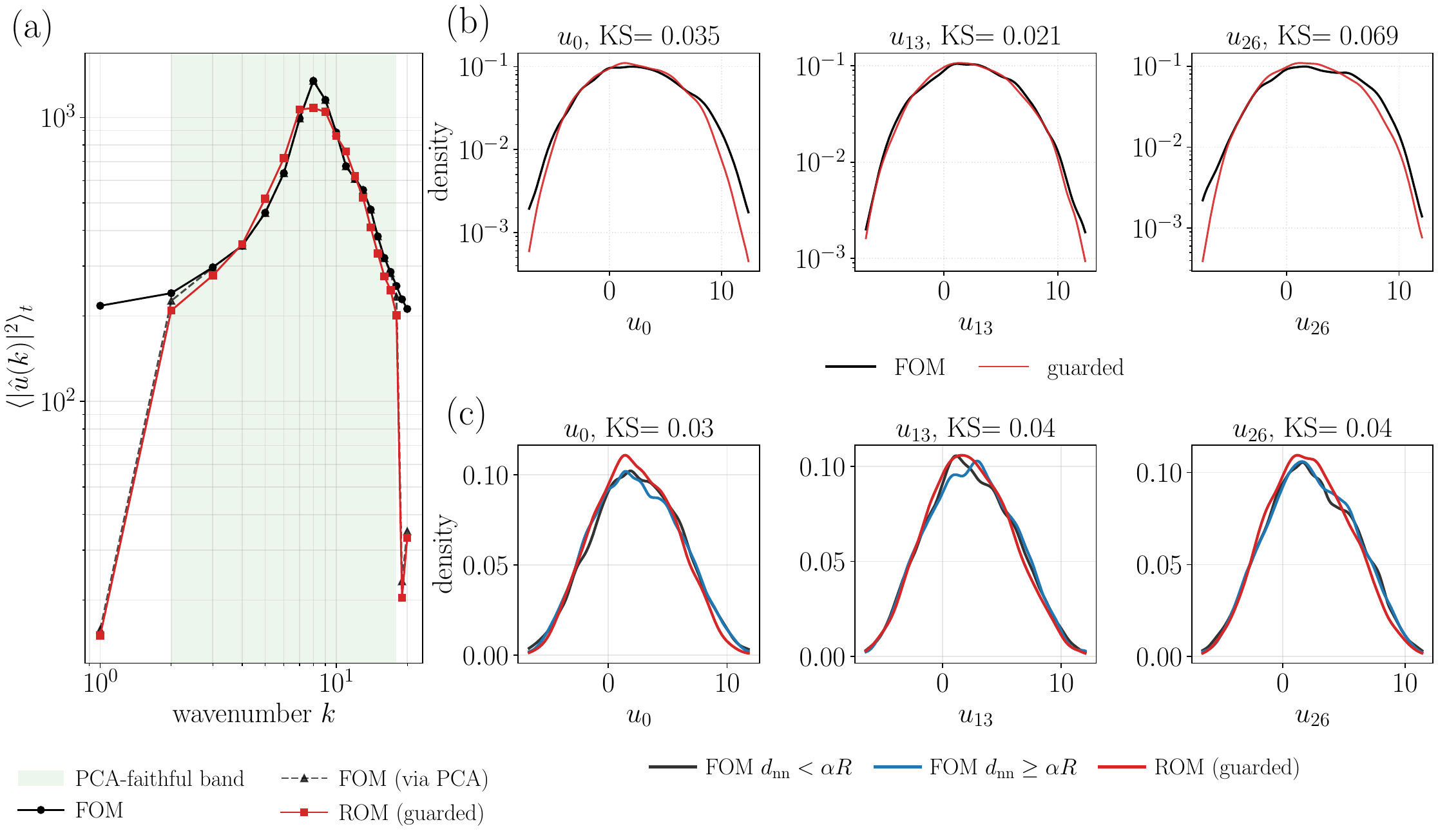}
	\caption{%
		Long-time statistics of the guarded model against the full-order reference. \emph{(a)} spatial energy spectrum, logarithmic deviation $0.044$ across the resolved band; \emph{(b)} per-mode marginal densities at three ring sites, Kolmogorov--Smirnov distance $0.033$ in the median; \emph{(c)} conditional densities at the same sites: the full-order state inside the trust region, in the tail beyond $\eta R_k$, and the guarded model, coinciding to within $0.04$.
	}
	\label{fig:app:invariant}
\end{figure}

The spatial energy spectrum of the guarded model follows the reference to a logarithmic deviation of
$0.044$ across the resolved band, peeling away only past the wavenumbers the
proper-orthogonal-decomposition truncation still resolves, and the per-mode marginal densities agree
to a Kolmogorov--Smirnov distance of $0.033$ in the median (\Cref{fig:app:invariant}a,~b).
Lorenz-96 is statistically homogeneous on the ring, so the three sites shown
stand in for all forty, and their near-identical shapes report that homogeneity
as much as the quality of the fit.
Two diagnostics not plotted here tell the same story: the two-point autocorrelation
time matches the reference at $0.30$, and the leading Lyapunov exponent is
recovered, the bare field returning $\Lambda_1 = 1.57$ against the reference
$1.68$ \citep{vlachas2020}.

Matching the global statistics leaves one loophole.
The trust radius is built from the same training distances the field is fitted to, so the corrector
might fire on legitimate but sparsely sampled on-attractor states, censoring the rare events the
full-order system visits, a distortion that a global comparison could average away.
Panel~(c) of \Cref{fig:app:invariant} rules this out.
At each site it sets three densities against one another: the full-order state
conditioned on lying inside the trust region, the full-order state conditioned on
lying in the tail beyond $\eta R_k$, and the guarded model.
The three are statistically indistinguishable, agreeing to a Kolmogorov--Smirnov
distance of at most $0.04$.
The tail the corrector acts on is dynamically the same as the bulk it leaves
alone, so the safeguard cannot be removing on-attractor structure: the structure
is identical in both bands.
The corrector confines the trajectory without moving its statistics.
\clearpage

\section{Supplementary diagnostics per test case}
\label{app:extra}

This appendix collects the per-test-case diagnostics that underpin the modelling
choices reported in the main text: the number of clusters $K$, the reduced
dimension retained by the proper-orthogonal-decomposition (POD) truncation, the
convergence of the RBF fit as the number of centres
grows, and the coarse energetic structure of the learned partition.
Lorenz-96 is shown at both forcings, $F = 8$ and $F = 10$, and in both the full
state and its SVD-reduced coordinates.
The Kuramoto--Sivashinsky and Kolmogorov cases additionally carry a reduction grid, since unlike
Lorenz-96 they are fields that must be projected onto a POD basis before the fit.
In every RBF-convergence figure the error metric is the normalised
root-mean-square error of the reduced-derivative prediction,
$\mathrm{NRMSE} = \mathrm{RMSE}_k / \sigma(\dot a_k)$, reported as the maximum,
median and minimum across the retained coordinates.

\subsection{Lorenz-96}
\label{app:extra:lor}

Lorenz-96 is already low-dimensional, so no projection precedes the fit.
The SVD-reduced coordinates are included only to show that the pipeline behaves the same way once a
reduction is imposed.
\begin{figure}[pos=H]
	\centering
	\includegraphics[width=\textwidth]{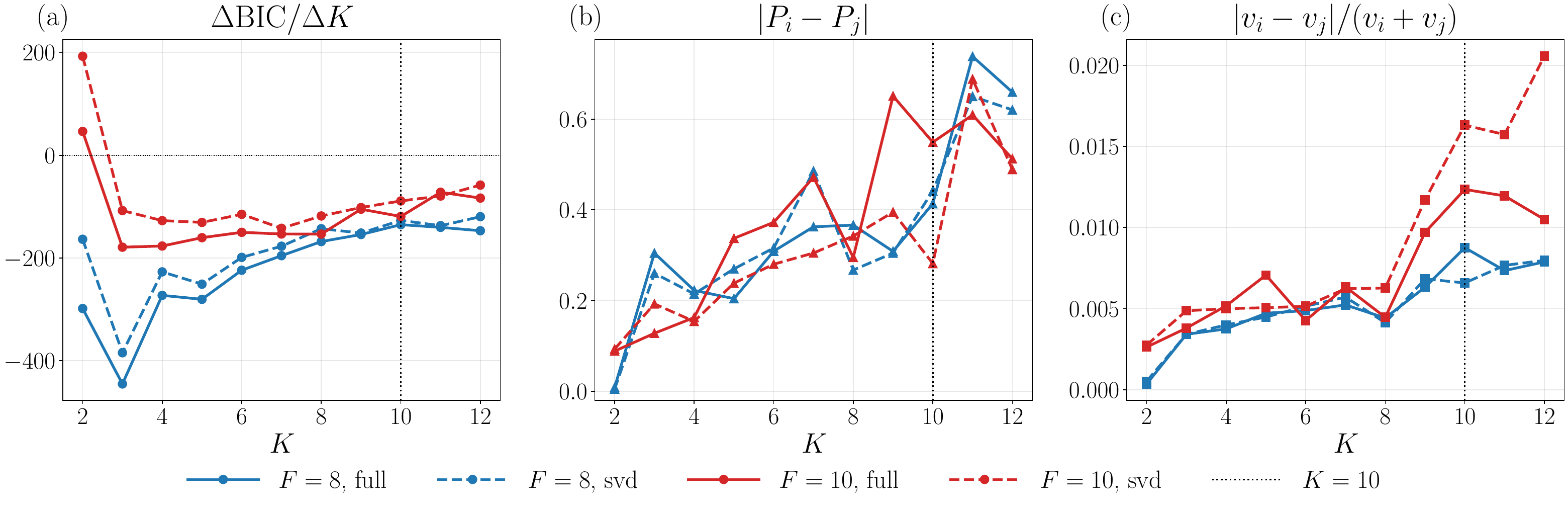}
	\caption{%
		Cluster-number selection for Lorenz-96, for the four combinations of $F \in \{8, 10\}$ and coordinate choice (full state versus SVD reduction); the vertical line marks $K = 10$. \emph{(a)} the marginal BIC gain $\Delta\mathrm{BIC}/\Delta K$, which carries no sharp elbow, so $K = 10$ is a deliberate choice. \emph{(b)} the elongation gap $|P_i - P_j|$ of adjacent clusters and \emph{(c)} the relative speed gap $|v_i - v_j|/(v_i + v_j)$, both growing with $K$.
	}
	\label{fig:app:lor:clusters}
\end{figure}
The marginal BIC gain carries no elbow at either forcing, so $K = 10$ is fixed by convention rather
than read from a knee (\Cref{fig:app:lor:clusters}a).
The elongation gap and the speed gap both climb steadily with $K$, so successive clusters
continue to separate rather than fragment a single basin (\Cref{fig:app:lor:clusters}b,~c).

\begin{figure}[pos=H]
	\centering
	\includegraphics[width=\textwidth]{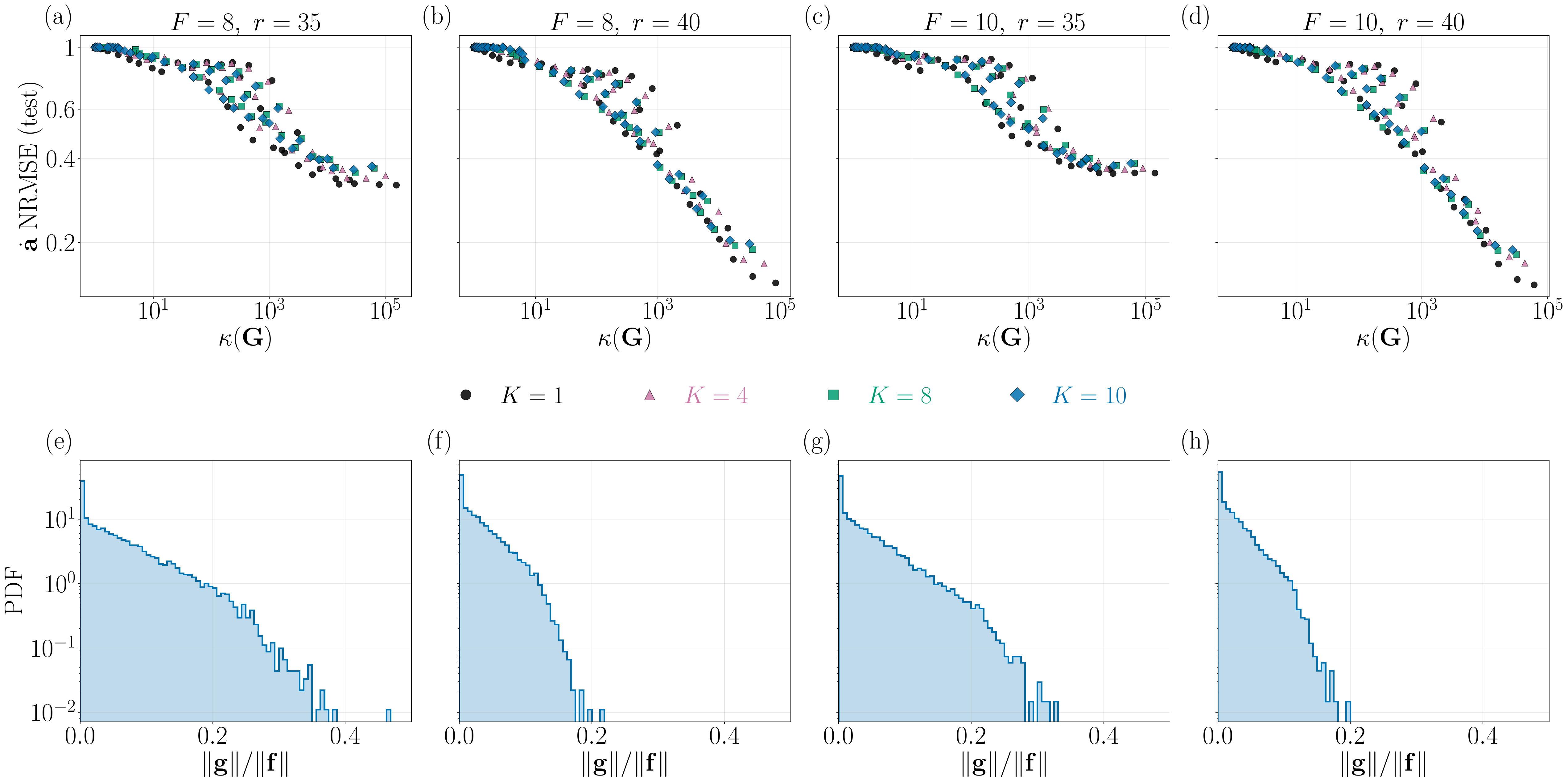}
	\caption{%
		Lorenz-96, one column per configuration ($F = 8$ and $F = 10$, SVD-reduced $r = 35$ and full $r = 40$ coordinates). \emph{(a--d)} test-set NRMSE of the reduced-derivative prediction against the condition number $\kappa(\mathbf G)$ of the Gram matrix, every $(\alpha, n_c)$ fit a point and one marker per number of clusters $K$; the error is a function of $\kappa$ alone. \emph{(e--h)} density of the corrector-to-field ratio $\|\mathbf g\|/\|\mathbf f\|$ along the long rollouts of the $K = 10$ model used in \cref{fig:l96invariant}; the spike at zero is the fraction of states where the corrector is off.
	}
	\label{fig:app:lor:rmse}
\end{figure}

The clusters resolve the phase of the travelling wave and the transitions advance it
(\Cref{fig:app:lor:energy}), more sharply at $F = 8$, where eight of the ten sit on the rim in phase
order, than at $F = 10$, where half fall towards the centre.
In the full coordinates the training and test errors of the fit are indistinguishable at every centre
count (\Cref{fig:app:lor:rmse}).
The SVD coordinates instead open a gap between the two and spread more widely, the reduced directions
being both harder to fit and slower to generalise.

\subsection{Kuramoto--Sivashinsky}
\label{app:extra:ks}

\begin{figure}[pos=H]
	\centering
	\includegraphics[width=\textwidth]{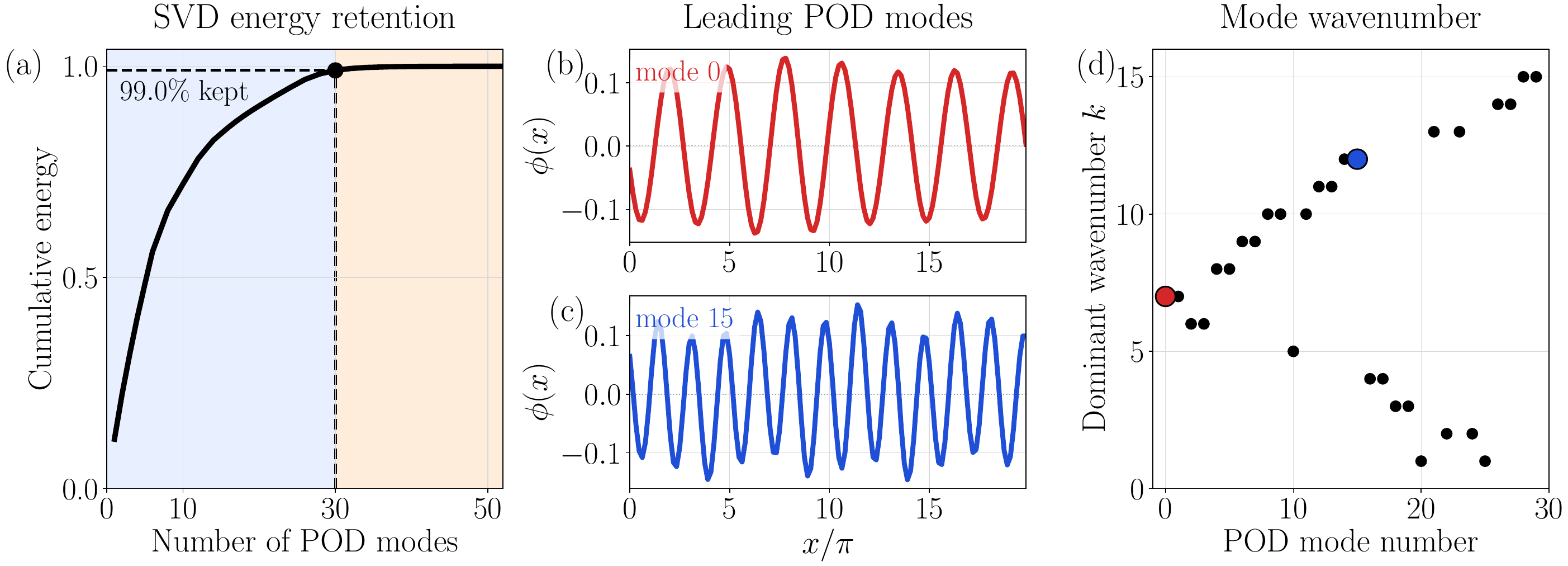}
	\caption{%
		POD truncation for Kuramoto--Sivashinsky. \emph{(a)} cumulative energy against the number of modes; $30$ modes retain $99.0\%$. \emph{(b, c)} two leading POD modes, clean harmonics: mode $0$ at $k \approx 7$, mode $15$ at $k \approx 12$. \emph{(d)} dominant wavenumber against mode index, a near-monotone ramp ordering the basis by scale.
	}
	\label{fig:app:ks:reduction}
\end{figure}

\begin{figure}[pos=H]
	\centering
	\includegraphics[width=\textwidth]{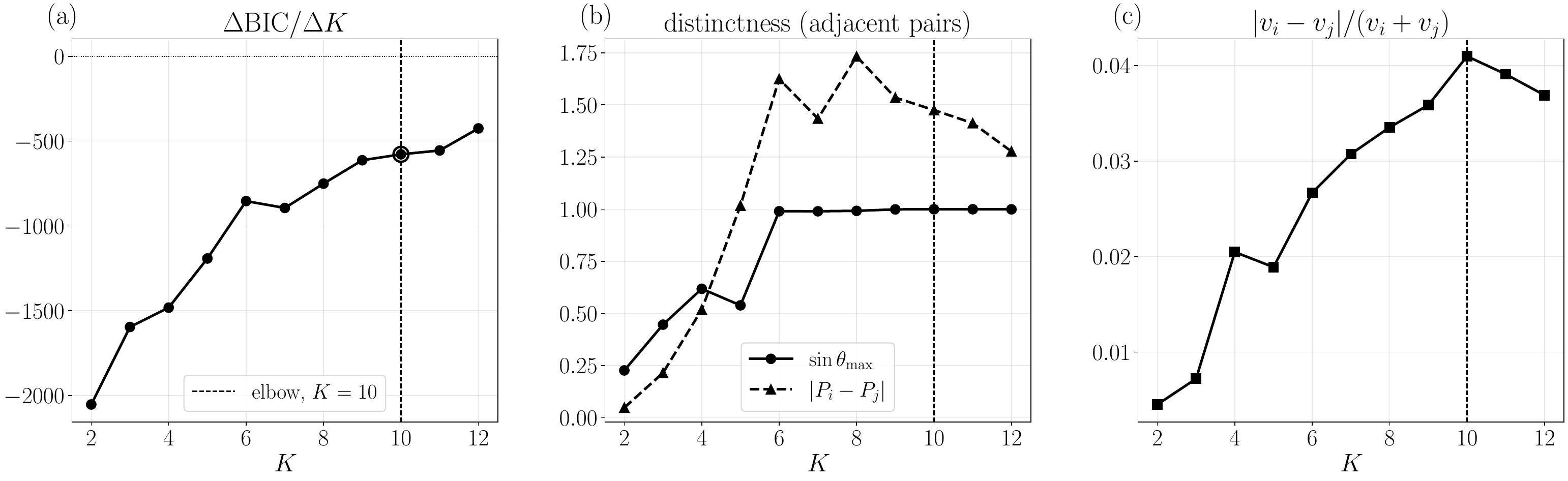}
	\caption{%
		Cluster-number selection for Kuramoto--Sivashinsky. \emph{(a)} the marginal BIC gain $\Delta\mathrm{BIC}/\Delta K$, elbow at $K = 10$ circled. \emph{(b)} adjacent-pair distinctness: the principal angle $\sin\theta_{\max}$ (solid) saturates to one from $K = 6$, the elongation gap $|P_i - P_j|$ (dashed) peaks in the same range. \emph{(c)} the relative speed gap $|v_i - v_j|/(v_i + v_j)$, maximal at $K = 10$.
	}
	\label{fig:app:ks:clusters}
\end{figure}

Thirty POD modes retain $99.0\%$ of the fluctuation energy, and the modes are clean harmonics
ordered by spatial scale, the dominant wavenumber rising almost monotonically with mode index
(\Cref{fig:app:ks:reduction}).
The BIC gain has a clear elbow at $K = 10$ (\Cref{fig:app:ks:clusters}a).
The principal angle between adjacent clusters saturates to one from $K = 6$ onward, so beyond that
point the clusters occupy genuinely distinct subspaces.
The resulting clusters fall into a low-energy six-cell family and a higher-energy seven-cell family,
with a short-residence group carrying the local seven-to-eight-cell defect
(\Cref{fig:app:ks:energy}).
The coexistence of the six- and seven-cell states is a genuine feature of the flow rather than an
artefact of the clustering.

\begin{figure}[pos=H]
	\centering
	\includegraphics[width=\textwidth]{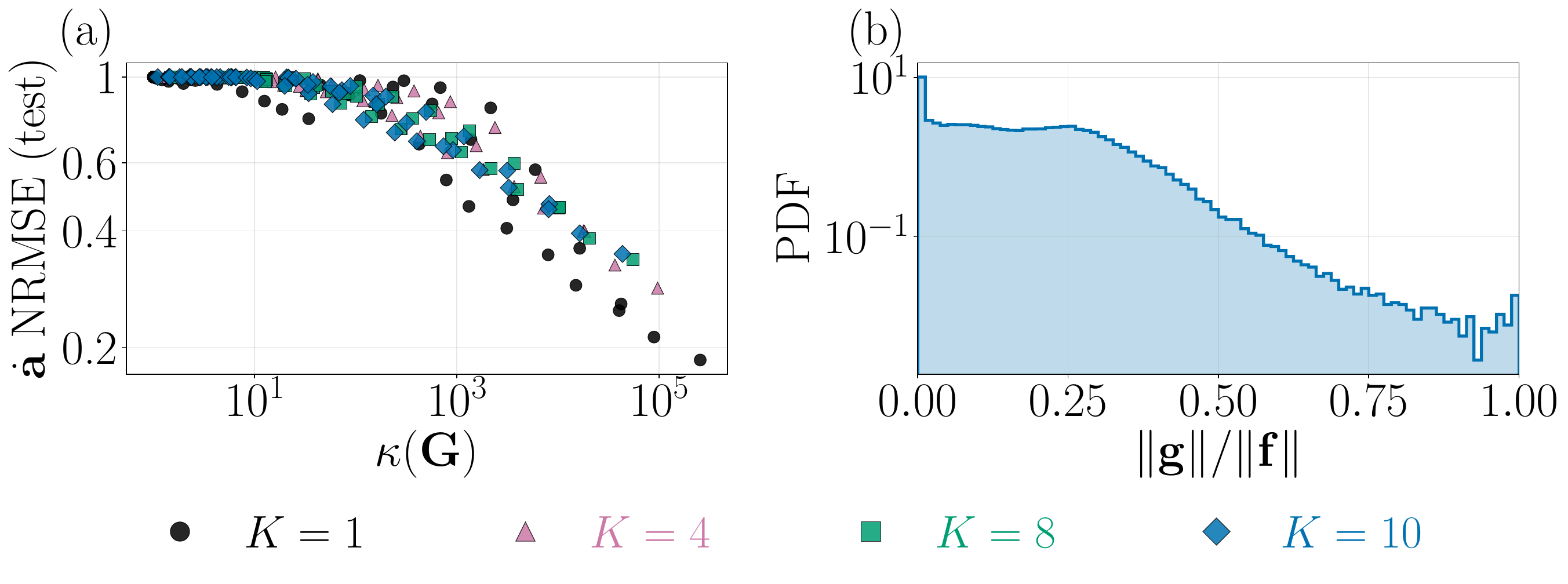}
	\caption{%
		Kuramoto--Sivashinsky. \emph{(a)} test-set NRMSE of the reduced-derivative prediction against the condition number $\kappa(\mathbf G)$ of the Gram matrix, every $(\alpha, n_c)$ fit a point and one marker per number of clusters $K$; the error is a function of $\kappa$ alone. \emph{(b)} density of the corrector-to-field ratio $\|\mathbf g\|/\|\mathbf f\|$ along $100$ rollouts of the calibrated $K = 10$ model, each of $100$ Lyapunov times; the spike at zero is the fraction of states where the corrector is dormant.
	}
	\label{fig:app:ks:rmse}
\end{figure}

The fit, \Cref{fig:app:ks:rmse}, collapses onto a single curve in $\kappa(\mathbf G)$ whatever the
number of clusters.
The corrector is active over most of the rollout, and its median pull is about a fifth of the learned
field, a larger share than the Lorenz-96 models ask for.

\subsection{Kolmogorov flow}
\label{app:extra:kol}
\begin{figure}[pos=H]
	\centering
	\includegraphics[width=\textwidth]{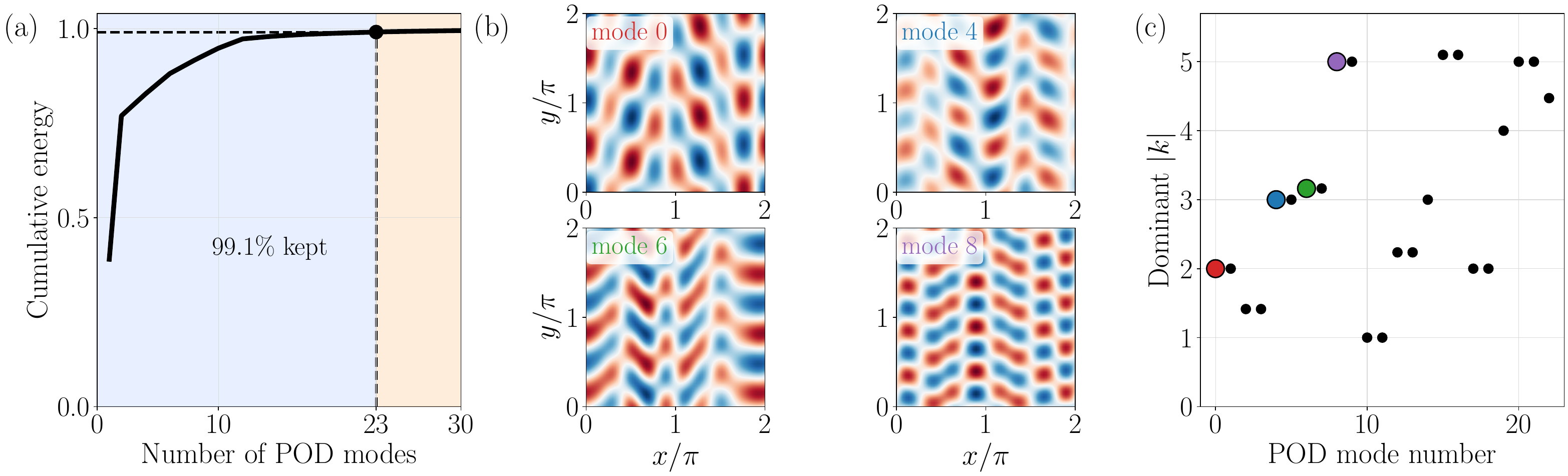}
	\caption{%
		POD truncation for Kolmogorov flow. \emph{(a)} cumulative energy against the number of modes; $23$ modes retain $99.1\%$. \emph{(b)} four leading POD modes, the characteristic Kolmogorov rolls. \emph{(c)} dominant wavenumber $|k|$ against mode index, rising broadly as finer scales enter the basis.
	}
	\label{fig:app:kol:reduction}
\end{figure}
Twenty-three POD modes retain $99.1\%$ of the fluctuation energy
(\Cref{fig:app:kol:reduction}).
The leading modes are the characteristic Kolmogorov rolls, and the dominant wavenumber broadens with
mode index.
\begin{figure}[pos=H]
	\centering
	\includegraphics[width=\textwidth]{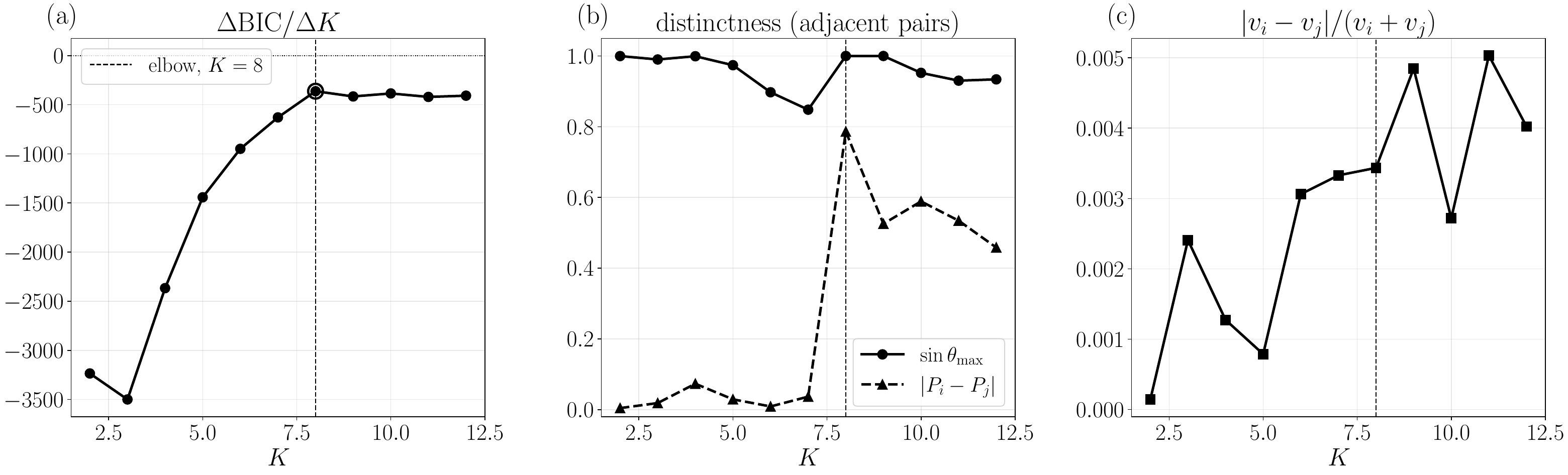}
	\caption{%
		Cluster-number selection for Kolmogorov flow. \emph{(a)} the marginal BIC gain $\Delta\mathrm{BIC}/\Delta K$, elbow at $K = 8$ circled. \emph{(b)} adjacent-pair distinctness: the principal angle $\sin\theta_{\max}$ (solid) stays near one throughout, while the elongation gap $|P_i - P_j|$ (dashed) jumps sharply at $K = 8$, where the clusters come to differ in elongation. \emph{(c)} the relative speed gap $|v_i - v_j|/(v_i + v_j)$.
	}
	\label{fig:app:kol:clusters}
\end{figure}
The BIC elbow falls at $K = 8$ (\Cref{fig:app:kol:clusters}).
Here the principal angle stays near one throughout, so it is the elongation gap
that discriminates: it jumps sharply at $K = 8$, so the elbow is exactly the point
at which adjacent clusters come to differ in elongation.

\begin{figure}[pos=H]
	\centering
	\includegraphics[width=\textwidth]{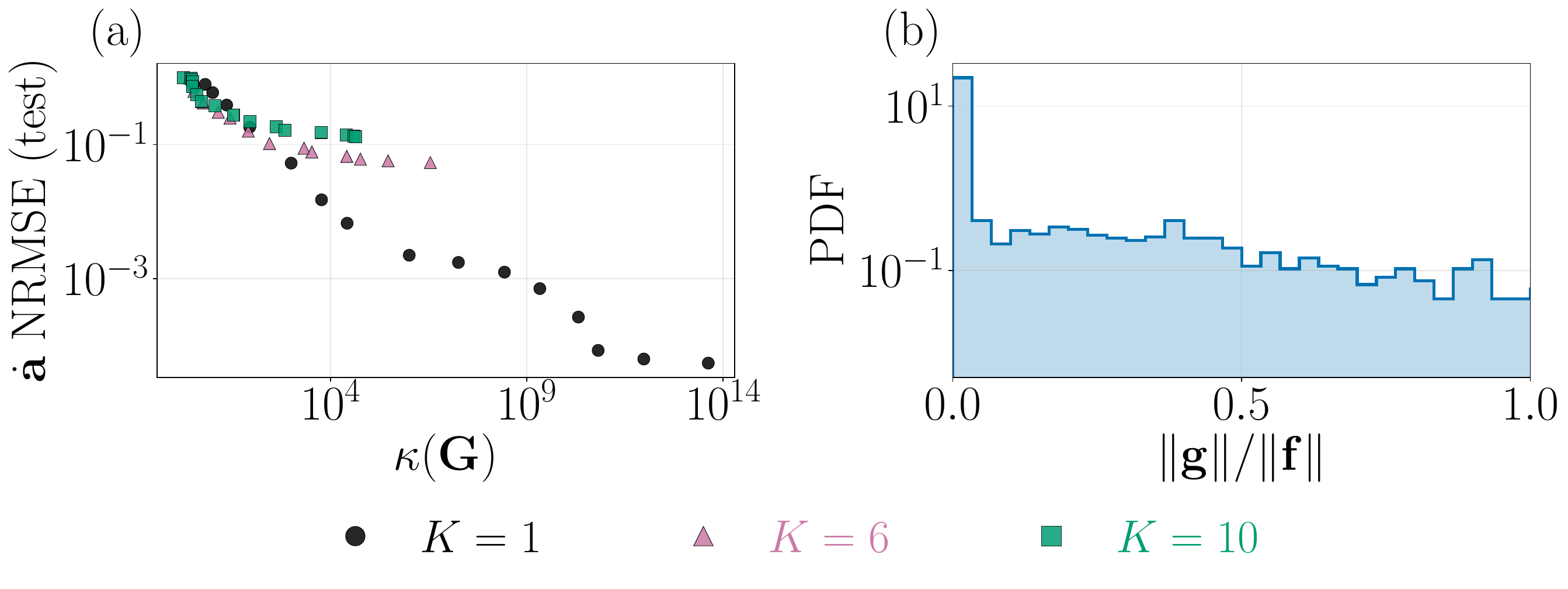}
	\caption{%
		Kolmogorov flow. \emph{(a)} test-set NRMSE of the reduced-derivative prediction against the condition number $\kappa(\mathbf G)$ of the Gram matrix, every $n_c$ fit a point and one marker per number of clusters $K$; unlike the chaotic cases the curves separate, a single cluster reaching $\sim\!10^{-4}$ where ten clusters floor two orders of magnitude higher. \emph{(b)} density of the corrector-to-field ratio $\|\mathbf g\|/\|\mathbf f\|$ along the rollout of \cref{fig:kol20}; the spike at zero is the fraction of states where the corrector is dormant, and the axis stops at unity, above which a further $9\%$ of the states fall.
	}
	\label{fig:app:kol:rmse}
\end{figure}

The partition tiles the phase of the torus, six clusters stepping one bin per transition and the
remaining two being a single region cut in two (\Cref{fig:app:kol:energy}).
The fit, \Cref{fig:app:kol:rmse}, reaches an NRMSE of order $10^{-4}$, and here the conditioning
alone does not set the error: at matched $\kappa(\mathbf G)$ a single cluster is three times more
accurate than ten, the torus being simple enough that partitioning it only constrains the kernels.
The corrector is dormant for three quarters of the rollout, but where it does act it is comparable to
the learned field, the price of the strength this case was given.

\clearpage
% Bibliography inlined below (compiled from references.bib with elsarticle-num-names) so the paper compiles without running bibtex.



\end{document}